\documentclass{jfm}
\usepackage{graphicx}
\usepackage{epstopdf, epsfig}
\usepackage{amsmath}
\usepackage{float}
\usepackage[dvipsnames]{xcolor}
\usepackage{natbib}
\usepackage{hyperref} 
\usepackage{bm}
\usepackage{lineno}
\usepackage{orcidlink}
\usepackage{svg}

\renewcommand{\d}		{\partial}

\newcommand{\dd}	    {{\rm d}}

\newcommand{\J}	        {\mathcal{J}}
\newcommand{\F}         {\mathcal{F}}
\newcommand{\complexconj}        {\text{c.c.}}
\newcommand{\A}         {\mathcal{A}}

\shorttitle{The increased drift of steep focusing waves}
\shortauthor{A. Blaser, L. Lenain, N. Pizzo}

\title{The increased drift of steep focusing surface gravity waves}

\author{Aidan Blaser\aff{1}\corresp{\email{ablaser@ucsd.edu}}, Luc Lenain\aff{1} and Nick Pizzo\aff{2}}

\affiliation{\aff{1}Scripps Institution of Oceanography, University of California San Diego,
La Jolla, CA 92037, USA \aff{2}Graduate School of Oceanography, University of Rhode Island, Narragansett, RI, USA 02881} 
\begin{document}
\maketitle

\begin{abstract}

Irrotational and monochromatic surface gravity waves possess a mean Lagrangian drift which transports mass and enhances mixing in the upper ocean. In the ocean, where many surface waves are present, it is commonly assumed that the mean Lagrangian drift can be computed independently for each wave component and summed. Here we show, using laboratory measurements and fully nonlinear simulations of steep focusing wave packets, that this assumption underpredicts the average transport in regions of wave focusing by up to $30\%$. To explain these enhancements, we derive a new exact method for constraining the local mean Lagrangian drift in general flows by working in the Lagrangian reference frame. From this method, we derive a higher-order expression for the local mean Lagrangian drift in narrow-banded wave fields governed by the nonlinear Schrödinger equation (NLSE) that predicts near-surface enhancements when waves focus and steepen. The theoretical predictions of the local transport agree with the experiments, particularly for smaller bandwidth packets where the NLSE approximation is most valid. These findings highlight that it is the local steepness of the wave field, not just the sum of the steepnesses of the linear (non-interacting) wave components, which sets the strength of these enhancements. 

\end{abstract}


\section{Introduction}

Irrotational surface gravity waves affect the transport of mass in the ocean through their mean Lagrangian drift \citep{Van2018}. For steady monochromatic plane waves, this drift is horizontally uniform and increases with wave steepness \citep{Stokes1847}. Ocean waves are neither steady nor monochromatic, and yet in most cases it is assumed that the total mean Lagrangian drift can be computed by treating the sea surface as a linear sum of non-interacting monochromatic plane waves \citep[e.g.,][]{kenyon1969}. In this paper we show that this assumption significantly underpredicts the near-surface mean Lagrangian drift when the surface becomes locally steep.

The mean Lagrangian drift impacts upper ocean processes across spatiotemporal scales greater than those of individual waves, making its accurate estimation crucial to a number of applications. For example, this drift directly influences the transport and dispersal of buoyant marine debris, such as plankton, plastics and oil spills \citep{vanSebille2020}. It is also widely understood that this vertically sheared Lagrangian mean flow interacts with the background vorticity field to tilt and stretch vortices, producing horizontal overturning cells indicative of Langmuir circulation \citep{Craik1976,Leibovich1983}. These overturning cells help mix the upper ocean, and many studies emphasize the need to parameterize these effects in large-scale models \citep[e.g.,][]{Belcher2012}. Any enhancements to the mean Lagrangian drift, especially in steep wave fields where its magnitude is largest, can therefore have a profound effect on these upper-ocean processes.

The impetus for this work came from a series of laboratory experiments \citep{lenain2019,sinnis21} which measured the total Lagrangian displacement of surface particles induced by breaking and non-breaking wave packets. These packets consisted of multiple wave components which were tuned to constructively interfere or focus at a prescribed location and time via dispersion. Wave breaking was found to greatly increase the Lagrangian transport, with the enhancements strongly dependent on each particle's distance from the breaking location. Interestingly, a similar, albeit weaker, spatial dependence was observed for steep non-breaking packets, with the largest enhancements occurring within the focusing region where the packet was most steep. This result was unexpected, since when viewed as a sum of linear monochromatic plane waves, the only differences between a focused and unfocused packet are relative phase shifts between wave components. If the total mean Lagrangian drift could be obtained by summing the individual drifts of each wave independently of the others, the relative phase shifts should be irrelevant. Thus, one should expect both the total drift and net transport to be spatially constant and independent of packet focusing. 

To supplement the limited laboratory data, we present numerical simulations of surface Lagrangian particle trajectories in equivalently defined packets using a fully nonlinear potential flow solver \citep{LHC1976,dold1992}. With a high spatial particle density, these simulations can better capture the spatial dependence of the surface Lagrangian transport. Repeating these simulations over a wide parameter space of steepness and bandwidth parameters reveals that the surface transport of particles averaged over the focusing region can exceed the spatially invariant predictions of linear theory by up to $30\%$. Some individual particles can even be transported up to twice this prediction, all without any wave breaking.

It should be clear that one cannot predict local enhancements to the mean Lagrangian drift without a local theory to explain it. By working in the Lagrangian reference frame, we derive a new exact technique for constraining the local mean Lagrangian drift of general wavy flows through the local mean pseudomomentum. This result is similar to the circulation theorem in generalized Lagrangian-mean (GLM) theory \citep{Andrews1978} but presented in a fully Lagrangian framework. Leveraging this new method, we derive a higher-order expression for the local mean Lagrangian drift in narrow-banded wave packets governed by the nonlinear Schrödinger equation \citep[NLSE,][]{Zakharov1968,Chu1970} which predicts near-surface enhancements when waves focus and steepen. We then use this analytical expression to estimate local enhancements to the mean Lagrangian transport for the simulations, and good agreement is found especially for lower bandwidths where the NLSE approximation is most valid.

This paper is organized as follows, in \S \ref{lintheory}, we introduce the equations of motion in the Lagrangian reference frame and derive a novel method to compute the local mean Lagrangian drift for general wavy flows. In \S \ref{datasection}, we define the focusing wave packets used, and numerically simulate their surface particle trajectories, comparing the results with laboratory data. In \S \ref{packets}, we derive the Lagrangian particle trajectories in narrow-banded waves and compute a higher order expression for the local mean Lagrangian drift, testing this theory against the simulations. In \S \ref{discussion}, we discuss the implications of these results in broader geophysical contexts.

\section{The mean Lagrangian drift of waves} \label{lintheory}

There are two natural coordinate systems for representing fluid motion within surface gravity waves: the Eulerian and the Lagrangian. The Eulerian frame solves for the fluid velocity as a function of fixed physical space and time and is mathematically appealing due to the fact that for two-dimensional irrotational and incompressible flow, the fluid velocity is analytic in the interior. This means that the entirety of the flow is determined by its behavior at the boundaries \citep{luke1967}. The fluid interior is governed by the linear Laplace equation
\begin{equation} \label{laplace}
    \nabla^2 \phi = 0 \, , \qquad \mathbf{u} = \nabla \phi \, ,
\end{equation}
for the velocity potential $\phi(x,y,t)$, whose spatial gradient is the Eulerian fluid velocity $\mathbf{u}(x,y,t)$. Despite the equation of motion being linear, the problem is made considerably more difficult due to the nonlinear boundary conditions \citep[see, for example, \S 3.1 of][]{Phillips1977},
\begin{align}
    \eta_t + \phi_x \eta_x - \phi_y = 0 \Big|_{y = \eta(x,t)} \,\label{KBC} , \\ \phi_t + \tfrac{1}{2}(\nabla \phi)^2 - g \eta = 0 \Big|_{y = \eta(x,t)} \, , \label{EulerianBC}
\end{align}
where $\eta(x,t)$ is the surface elevation, an introduced independent variable not known \emph{a priori}, and $g$ is the acceleration due to gravity, which in our notation points in the $-\hat{y}$ direction, with subscripts indicating partial derivatives. While it is common to evaluate these boundary conditions by expanding in a Taylor series about the still water level $y=0$, this introduces infinitely many nonlinear terms which are in practice truncated by invoking some small parameter which is typically related to the surface wave slope. To compute the physical trajectories of fluid particles as functions of their initial positions and time, one must then integrate the coupled pathline equations
\begin{equation} \label{pathline}
    \frac{\dd x}{\dd t} = u = \phi_x (x,y,t) \, , \qquad \frac{\dd y}{\dd t} = v = \phi_y(x,y,t) \, ,
\end{equation}
holding particles fixed. It was by this method that \cite{Stokes1847} first computed the magnitude and profile of the mean Lagrangian drift for irrotational, monochromatic waves, though the complexity of integrating the nonlinear pathline equations limited the accuracy of his solution to second order in wave slope. If the desired result is to compute the mean Lagrangian motion of particles, it is much more natural to work directly in the Lagrangian reference frame, where the physical particle trajectories $(x,y)$ are explicitly solved for as functions of general labeling coordinates $(\alpha,\beta)$ and time $\tau$, which we distinguish from the usual notation $t$ to emphasize that a partial derivative with respect to $\tau$ holds particle labels fixed (i.e., equivalent to the material derivative $\frac{\dd}{\dd t}$ in the Eulerian frame). One can view these physical trajectories as a time-dependent mapping from a certain ``label space" to physical space with a corresponding Jacobian determinant
\begin{equation}\label{Jacobian}
    \J \equiv \frac{\d (x,y)}{\d(\alpha,\beta)}= x_\alpha y_\beta - x_\beta y_\alpha \, ,
\end{equation}
whose value determines how infinitesimal areas are scaled by the nonlinear mapping. Since incompressible flow requires that a small collection of particles $\dd \alpha \, \dd \beta$ enclose the same physical area  $\J^{-1}\dd x \, \dd y$ as the flow evolves, we see that $\J$ must be everywhere time independent and nonzero,
\begin{equation}
    \J_\tau = x_{\alpha \tau}y_\beta + x_\alpha y_{\beta \tau} - x_{\beta\tau}y_\alpha - x_\beta y_{\alpha\tau} = 0 \, , \label{incompressibility}
\end{equation}
The particular choice of labeling particles must not affect the dynamics and thus represents an important gauge freedom in fluid mechanics \citep{Salmon2020}. For simplicity we hereinafter choose to work with a labeling gauge such that $\J = 1$, so that areas in label space equal areas in physical space. While we can still define a velocity potential in the Lagrangian frame for irrotational flow, the generally nonlinear mapping between physical and label space implies that the form of the Laplacian operator is more complicated in label space as these maps are not generally harmonic. Instead, we turn to the full Euler equations which in the Lagrangian frame are written as \cite[][Art. 15]{LAMB1932}
\begin{equation}\label{xEuler}
   \J x_{\tau\tau} + p_\alpha y_\beta - p_\beta y_\alpha = 0 \, ,
\end{equation}
\begin{equation}\label{yEuler}
    \J y_{\tau\tau} + p_\beta x_\alpha - p_\alpha x_\beta + \J g = 0 \, ,
\end{equation}
where $p$ is the fluid pressure. Note that while the material acceleration is greatly simplified in the Lagrangian frame, the pressure gradient force is no longer represented by a simple linear operator. In practice any Eulerian quantity or operator can be converted to the Lagrangian frame through the Jacobian. For example, the vorticity of the fluid can be converted to the Lagrangian frame via the following steps
\begin{equation}\label{vorticity}
    q = v_x - u_y = \frac{\d (v,y)}{\d(x,y)} + \frac{\d (u,x)}{\d (x,y)} = \frac{1}{\J} \bigg(\frac{\d(y_\tau,y)}{\d (\alpha,\beta)} + \frac{\d (x_\tau,x)}{\d (\alpha,\beta)}\bigg)\, ,
\end{equation}
%
where $q$ is conserved on particles (i.e., $q_\tau = 0)$ for two-dimensional inviscid flow, which can be seen by eliminating $p$ between the two Euler equations. The strict condition of irrotational flow thus imposes the following constraint on the fluid trajectories
\begin{equation}\label{irrotational}
    \J q = x_{\tau\alpha}x_\beta - x_{\tau\beta}x_\alpha + y_{\tau \alpha} y_\beta - y_{\tau\beta}y_\alpha = 0 \, .
\end{equation}
To close the system, we impose the following boundary conditions; first, that the pressure vanishes up to a constant at the free surface which we label by our choice as $\beta=0$,
\begin{equation}\label{pboundary}
    p(\beta=0) = 0 \, ,
\end{equation}
and second, that the vertical velocity vanishes approaching the bottom at infinite depth,
\begin{equation}\label{ytboundary}
    y_\tau(\beta \rightarrow -\infty) = 0 \, .
\end{equation}

Note that while we have necessarily abandoned the simplicity of Laplace's equation for more complicated nonlinear equations of motion \eqref{xEuler}--\eqref{yEuler}, what we have gained from this approach is having simple boundary conditions without potentially infinite nonlinear terms which necessitate small amplitude approximations. In addition, as vorticity is conserved on particles, adding arbitrary vorticity to particles is straightforward in the Lagrangian frame as opposed to in the Eulerian frame where Laplace's equation would have to be replaced with the full nonlinear Euler equations alongside the nonlinear boundary conditions. 

\subsection{Mean Lagrangian drift of general flows}

While directly solving the Euler equations \eqref{xEuler}--\eqref{yEuler} subject to $\J=1$, $q=0$ and the boundary conditions \eqref{pboundary}--\eqref{ytboundary} will yield particle trajectories that explicitly contain the mean Lagrangian drift, this offers little physical insight into its origin. Previous studies connected the mean Lagrangian drift, or equivalently the mean Lagrangian momentum density, to other physical quantities such as vorticity and energy \citep{Pizzo2023,blaser_momentum_2024}, but these results necessarily assumed waves that were steady and monochromatic. In this section, we introduce a new method of constraining the mean Lagrangian drift for completely general flows. To do so, we start by considering the circulation of a material loop $\mathcal{C}$, which is defined as
\begin{equation}
    \Gamma \equiv \oint_{\mathcal{C}} \bm{u} \cdot \dd \bm{\ell} = \iint_{\Omega} (\bm{\nabla} \times \bm{u}) \cdot \bm{\hat{n}} \, \dd x \, \dd y \, ,
\end{equation}
with the last relation due to Stokes' theorem for the area $\Omega$ enclosed by the contour where $\bm{\hat{n}}$ is the unit outward normal. We simplify here to two-dimensional flow, but the following results may be readily extended to three-dimensions \citep{Salmon1988}. Just as with the vorticity, we can rewrite the circulation in Lagrangian coordinates via the chain rule,
\begin{equation}
    \Gamma = \oint_{\mathcal{C}_0} \Big(x_\tau \bm{\nabla_\alpha}x + y_\tau \bm{\nabla_\alpha}y\Big) \, \cdot \dd \bm{\alpha} = \iint_{\Omega_0} \J q \, \dd \alpha \, \dd \beta \, ,
\end{equation}
where $\bm{\nabla_\alpha} = (\d_\alpha,\d_\beta)$ is the gradient operator in label space. The contour $\mathcal{C}_0$ is now a contour in label space and is therefore fixed in time by definition; the same goes for the enclosed area $\Omega_0$. If we decompose the Lagrangian trajectories into an initial location and deviation,
\begin{equation}
    x = \alpha + \xi(\alpha,\beta,\tau) \, , \qquad y = \beta + \zeta(\alpha,\beta,\tau) \, ,
\end{equation}
so that $\alpha$ and $\beta$ can be seen as ``horizontal" and ``vertical" labels respectively, we can rewrite the circulation as
\begin{equation} \label{pseudomomentum}
    \Gamma = \oint_{\mathcal{C}_0} (\bm{x}_\tau - \bm{\mathrm{p}}) \cdot \dd \bm{\alpha} \, ,
\end{equation}
where $\bm{x}_\tau = (\xi_\tau,\zeta_\tau)$ is the Lagrangian velocity, and $\bm{\mathrm{p}} =-(\xi_\tau \xi_\alpha + \zeta_\tau \zeta_\alpha, \, \xi_\tau\xi_\beta + \zeta_\tau \zeta_\beta) $ is identified as the Lagrangian pseudomomentum. While its form looks identical to pseudomomentum as defined in generalized Lagrangian-mean (GLM) theory \citep{Andrews1978,Buhler2014}, they are still distinct since the displacement vector in GLM is a function of the Lagrangian mean trajectory, not Lagrangian particle labels. For irrotational flows where $\Gamma=0$ for all closed loops, \eqref{pseudomomentum} implies that the label space curl of the velocity must be everywhere equal to the label space curl of the pseudomomentum,
\begin{equation} \label{fullpseudo}
    \bm{\nabla}_{\bm{\alpha}} \times \bm{x}_\tau = \bm{\nabla}_{\bm{\alpha}}\times \bm{\mathrm{p}} \, ,
\end{equation}
analogous to the celebrated result in GLM \citep[][Ch. 10]{Buhler2014}. Since what we are interested in is the mean component of the velocity, we can take an average of \eqref{fullpseudo} to get
\begin{equation} \label{generalDrift}
    \bm{\nabla}_{\bm{\alpha}} \times  \langle\bm{x}_\tau \rangle = \bm{\nabla}_{\bm{\alpha}} \times \langle \bm{\mathrm{p}} \rangle \, ,
\end{equation}
where the angle brackets represent any general averaging operator that commutes with the curl, such as a time mean or convolutional average. It is worth pausing here for a moment to unpack this result, which states that for irrotational flow, the curl of the mean Lagrangian drift is exactly set by the curl of the mean pseudomomentum so that any modification to one immediately affects the other. Viewing the mean Lagrangian drift as essentially tethered to the mean pseudomomentum highlights its role as not simply a passive byproduct of the waves, but as a dynamic mean flow in its own right. This view will be especially helpful when we turn to the mean Lagrangian drift of narrow-banded wave packets. However, for completeness, we will use this new general framework to compute the mean Lagrangian drift for linear waves in the following subsections.

\subsection{Monochromatic waves}

We start with the classical example of a linear deep-water monochromatic wave with wavenumber $k$ and constant amplitude $A$ where the nondimensional steepness $Ak$ is assumed to be small. Following the method of \cite{Salmon2020}, Ch. 1, we assume a wavelike solution for $x$ and $y$ after expanding about a hydrostatic state of rest ($x=\alpha$, $y=\beta$, $p = -g\beta$),
\begin{align} \label{monoX}
    x(\alpha,\beta,\tau) &= \alpha - A e^{k \beta} \sin(k\alpha - \omega \tau) \, , \\
    y(\alpha,\beta,\tau) &= \beta + Ae^{k \beta}\cos(k\alpha - \omega \tau) \, , \label{monoY} \\ 
    p(\beta) &= -g \beta \, , \label{monoP}
\end{align}
where $\omega^2 = g k$ is the linear deep-water dispersion relation determined by substituting \eqref{monoX}--\eqref{monoP} into the Euler equations \eqref{xEuler}--\eqref{yEuler}. These simple circular trajectories are in fact exact solutions to the Euler equations known as \cite{Gerstner1802} waves. However, these waves are not irrotational, which can be seen by computing their vorticity using \eqref{vorticity}. From \eqref{generalDrift}, we see that irrotational flow requires that the curl of the mean Lagrangian drift be equal to the curl of the mean pseudomomentum. Computing the pseudomomentum of \eqref{monoX}--\eqref{monoY} yields only a horizontal component
\begin{equation}
    \mathrm{p} = -(\xi_\tau \xi_\alpha + \zeta_\tau\zeta_\alpha) = A^2 k \omega e^{2 k \beta} \, ,
\end{equation}
that varies only with depth. Taking the mean to be a long time average following a fixed particle, from \eqref{generalDrift} we require
\begin{equation}
    \frac{\d \langle y_\tau \rangle}{\d \alpha} - \frac{\d \langle x_\tau \rangle}{\d \beta} = -\frac{\d \langle \mathrm{p}\rangle}{\d \beta} = -2 A^2 k^2 \omega e^{2 k \beta} \, . \label{monochromaticcurl}
\end{equation}
On physical grounds we can assume there is no mean vertical motion, so that the solution to \eqref{monochromaticcurl} is
\begin{equation}
    \langle x_\tau \rangle = A^2 k \omega e^{2 k \beta}\, ,
\end{equation}
where the arbitrary constant can be removed in the frame where the velocity of fluid at depth vanishes. This the classical Stokes drift. We reproduce it here as an example of our general method but also because it shows how the second-order mean flow is constrained by first order orbital motion, due to the pseudomomentum being a quadratic quantity. This carries to higher order corrections as well; since the particle displacements in surface gravity wave fields $(\xi,\zeta)$ are always first order quantities or higher, one needs only to constrain trajectories valid to order $n$ to constrain the drift to order $n+1$.

\subsection{Multiple waves -- linear theory}

Following \cite{Pierson1961}, if our initial conditions instead consist of a discrete spectrum of $N$ deep-water plane waves traveling in the same direction, to first order we have
\begin{align}
    x &= \alpha - \sum_{n=1}^N A_n e^{k_n \beta} \sin(k_n\alpha - \omega_n \tau + \varphi_n) \, , \\
    y &= \beta + \sum_{n=1}^N A_n e^{k_n \beta} \cos(k_n \alpha - \omega_n \tau + \varphi_n)\, , \\
\end{align}
where $A_n$, $k_n$, $\omega_n$ and $\varphi_n$ are the amplitude, wavenumber, frequency and arbitrary initial phase of each wave component respectively. It is assumed that each wave's steepness $A_n k_n$ is small and, importantly for this analysis, constant. The horizontal component of the pseudomomentum is given by products of sums, but taking the mean to be a long time average following a fixed particle, we have
\begin{equation} \label{pseudoPierson}
    \langle\mathrm{p}\rangle =-\langle\xi_\tau \xi_\alpha + \zeta_\tau\zeta_\alpha\rangle= \sum_{n=1}^N A_n^2 k_n \omega_n e^{2 k_n \beta} \, ,
\end{equation}
where any cross terms vanish in the time mean due to to the fact that $A_n$, $k_n$, $\omega_n$ and $\varphi_n$ are constant. From \eqref{generalDrift},
\begin{equation}
    \frac{\d \langle y_\tau \rangle}{\d \alpha} - \frac{\d \langle x_\tau \rangle}{\d \beta} = -\frac{\d \langle \mathrm{p} \rangle}{\d \beta} \ = -\sum_{n=1}^N 2A_n^2 k_n^2 \omega_n e^{2 k_n \beta} \, .
\end{equation}
Once again, the only physically valid solution is balanced by $\langle x_\tau \rangle$, so that
\begin{equation}\label{piersondrift}
    \langle x_\tau \rangle = \sum_{n=1}^N A_n^2 k_n \omega_n e^{2 k_n \beta} \, ,
\end{equation}
where the constant of integration vanishes in the frame where the fluid interior is at rest. We see that according to lowest order theory, the total mean Lagrangian drift for a linear wave field is a simple sum of the individual drifts of each wave component. While the full second-order particle trajectory solutions contain bounded second order harmonics \citep{Pierson1961,nouguier2015} which can be interpreted as local fluctuations to the mean Lagrangian drift, these terms are fully oscillatory and do not contribute to the long time transport regardless of how the initial phases $\varphi_n$ are tuned. From this theory, the effect of local steepness fluctuations to the mean Lagrangian drift is symmetric; any local increases during constructive interference are canceled by local decreases during destructive interference. We would therefore expect the total transport of a passing wave packet, expressed as a sum of plane waves, to be similarly invariant to local wave focusing. In the following section, we investigate the Lagrangian transport of focusing wave packets, presenting numerically simulated particle trajectories alongside laboratory data.

\section{Lagrangian transport due to focusing wave packets} \label{datasection}

We now narrow our scope to that of spatially compact focusing wave packets. First we define these packets and provide a linear prediction of their induced surface Lagrangian transport. Next, we introduce the fully nonlinear solver used to simulate the Lagrangian trajectories of surface particles, and show the results of these simulations for a range of packet parameter space. Finally, we compare the results of the simulations and laboratory experiments against the predictions of linear theory.

\subsection{Packet initialization}

We define our packets as in \cite{Rapp1990, Drazen2008,sinnis21} to focus according to linear theory at a prescribed space and time
\begin{equation}\label{focusEta}
    \eta(x,t) = \sum_{n=1}^N A_n \cos(k_n (x - x_f) - \omega_n (t - t_f)) \, ,
\end{equation}
where $\eta(x,t)$ is the Eulerian free surface displacement, $A_n$ is the amplitude of each discrete wave, $k_n$ and $\omega_n$ represent the respective wavenumber and frequency of each component, both positive as all wave components travel to the right, and $x_f$ and $t_f$ denote the focusing location and time respectively according to linear theory. We consider a uniformly distributed spectrum in frequency space, so that our frequencies can be expressed as

\begin{equation} \label{omegadef}
    \omega_n = \omega_c (1 + \Delta (\tfrac{n}{N} - \tfrac{1}{2})) \, ,
\end{equation}
where $\omega_c$ is the central frequency (so that $k_c = \omega_c^2/g$ is the central wavenumber) and $\Delta$ is the non-dimensional bandwidth which sets the time and space scales of the focusing event and must be less than $2$ to ensure positive frequencies. In addition, as the slope of waves is an indicator of their nonlinearity, we wish to define the amplitudes $A_n$ such that at focusing, the linear prediction of the maximum slope equals some prescribed value $S$. Therefore, we define
\begin{equation}
    S = \sum_{n=1}^N A_n k_n \, .
\end{equation}
Thus, given the linear deep-water dispersion relationship $\omega_n^2 = g k_n$, we can determine the values of $A_n$, assuming the slope of each mode is equal following \cite{Drazen2008} (i.e., $A_n = S/(Nk_n)$). Placed in this formulation, the wave packets we consider are primarily functions of two independent variables, $S$ and $\Delta$, which will be used as our parameter space.

The linear prediction of the surface mean Lagrangian drift is given by a simple sum of the lowest order contributions \eqref{piersondrift}
\begin{equation}\label{lineardrift}
    \langle x_\tau \rangle\Big|_{\beta=0} = \sum_n^N A_n^2 k_n \omega_n = \sum_n^N \frac{S^2}{N^2} \frac{\omega_n}{k_n} \, .
\end{equation}
Based solely on \eqref{lineardrift}, the surface mean Lagrangian drift scales as $\frac{1}{N}$, which implies that a packet with more waves experiences less drift, despite the fact that $N$ simply represents the spectral resolution of the packet, whose form converges as $N \rightarrow \infty$. This is due to the fact that the temporal periodicity of \eqref{focusEta} is given by
\begin{equation}\label{Tperiodicity}
    T_p = \frac{2\pi N}{\omega_c \Delta} \, ,
\end{equation}
so that as $N$ increases, the time between subsequent packets also increases. Since what we are after is not the mean Lagrangian drift itself, which according to \eqref{piersondrift} is the same for all particles at all times since it treats the packet as a sum of monochromatic plane waves, we instead compute the total linear surface Lagrangian transport $\delta x_{\text{lin}}$ after a single packet has passed. This is done by integrating \eqref{lineardrift} in time over the temporal periodicity of the packet \eqref{Tperiodicity},
\begin{equation}\label{lineartransport}
    \delta x_\text{lin} = \int_0^{T_p} \langle x_\tau \rangle \, \dd\tau \Big|_{\beta=0}= \langle x_\tau \rangle T_p \Big|_{\beta=0}= \frac{2\pi S^2}{\omega_c \Delta} \bigg(\frac{1}{N}\sum_n^N \frac{\omega_n}{k_n}\bigg)\, .
\end{equation}
We see that the linear prediction of the total surface Lagrangian transport should scale with $S^2$ as one should expect for the lowest order theory. The transport scaling inversely with $\Delta$ should also be expected as the packet width in physical space is inversely proportional to its width in wavenumber space via the generalized uncertainty principle \citep{sinnis21}. The last term represents a spectrally weighted phase speed. For $N$ large, \eqref{lineartransport} can be approximated in closed form from \eqref{omegadef} as
\begin{equation}\label{transportapprox}
    \delta x_\text{lin} \approx \frac{2 \pi S^2}{k_c} f(\Delta) \, , \qquad f(\Delta) = \frac{1}{\Delta^2}\ln\bigg(\frac{1 + \Delta/2}{1 - \Delta/2}\bigg) \, ,
\end{equation}
where $f(\Delta)$ represents the linear bandwidth dependence on the total transport, found by approximating the sum in \eqref{lineartransport} as an integral. While this full expression is slightly more complicated than the heuristic argument given above, $f(\Delta)$ is well approximated by $\Delta^{-1}$ when $\Delta$ is small. 

\subsection{Numerical simulations of Lagrangian trajectories}

To simulate the Lagrangian trajectories of surface particles within these packets, we employ a fully nonlinear mixed Eulerian-Lagrangian potential flow solver \citep{dold1992}. Originally developed by \citep{LHC1976}, this method takes advantage of the fact that at a fixed time, the Eulerian and Lagrangian velocities are equal since a particle occupies a single fixed location at a fixed time. Because solutions to Laplace's equation \eqref{laplace} are uniquely determined by the boundary conditions, only the surface needs to be simulated, assuming a constant or infinite depth and a periodic domain. By initializing Lagrangian particles with initial positions $(x_0,y_0)$ and velocity potential $\phi_0$, this solver computes the gradient of $\phi$ at the surface given its value via Cauchy's integral theorem at each time step. This allows for the particle positions $(x,y)$ to evolve via the pathline equations \eqref{pathline}, with the velocity potential evolving according to Bernoulli's equation at the free surface \eqref{EulerianBC}. Because Lagrangian particles naturally cluster at wave crests where the spatial curvature is strongest, the resolution of this method is naturally adaptive, and numerous studies \citep{dommermuth1988,skyner_comparison_1996} have validated the accuracy and validity of this numerical method.

For our simulations, we chose a central wave frequency of $1$ Hz, so that $\omega_c = 2\pi$ rad/$s$ and $k_c= (2\pi)^2/g$ rad/$m$. The domain length was chosen to be $L = 150$ m, long enough so that the entire packet could fully pass over a large enough collection of particles to obtain an unambiguous measure of total Lagrangian transport before any signal wrapped around due to the periodicity of the domain. To fully resolve the free surface, we systematically increased the number of Lagrangian particles used until convergence was reached at 2048 particles, or around 20 per central wavelength. The depth of the water is taken to be infinitely deep, and the packets were initialized to start $20 \, m$ away from the prescribed focusing location so that there were sufficient particles within the focusing region that both started and ended at rest. We defined the linear prediction of the focusing time as $t_f = x_f/c_g$, where $c_g = \tfrac{1}{2}\sqrt{g/k_c}$ is the central group velocity according to linear theory. Lastly, we chose to use $N=1000$ wave modes so that the spectral resolution is sufficiently high to converge the physical shape of the packet.

The procedure for simulating these packets is as follows. First, we initialize the horizontal positions $x_0$ to be evenly spaced along the domain. Then, using \eqref{focusEta}, the vertical initial positions $y_0$ are found for prescribed values of $S$ and $\Delta$. To ensure only one packet is used and that the domain is totally periodic, a windowing function is applied to $y_0$ with minimal energy loss (less than one part in 100). The initial velocity potential is found according to linear theory by performing a Fourier transform on the windowed $y_0$, multiplying each Fourier amplitude by $\omega_n/k_n$, and performing an inverse transform. The simulations were repeated for a parameter space spanning $\Delta = 0.3$, incremented by $0.1$ until $\Delta = 1.3$, and $S = 0.05$, incremented by $0.02$ until the packet broke, which agreed well with the results of \cite{pizzo2021b}, who numerically investigated the breaking threshold $S_*$ of these same packets as a function of bandwidth. To improve parameter space resolution near $S_*$, we ran additional simulations near this threshold.

\begin{figure}
    \centering
    \includegraphics[width=1.0\textwidth]{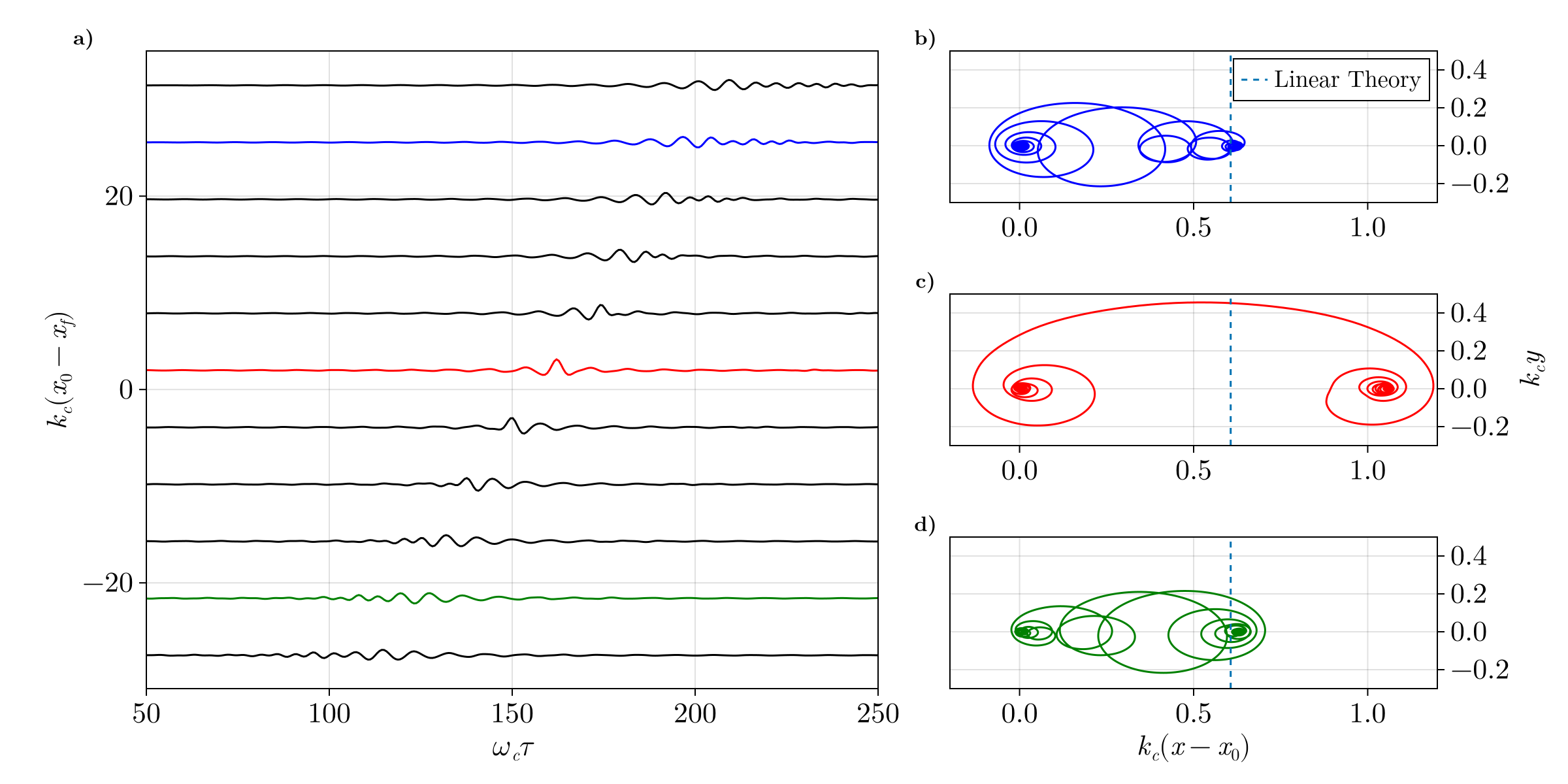}
    \caption{Surface particle trajectories in a focusing wave packet with $\Delta = 0.8$ and $S = 0.27$. In panel $(a)$, the vertical elevation of particles is shown in time as a function of their initial location from the linear prediction of focusing $x_f$, normalized by the central wavenumber $k_c$. The colored lines represent particles downstream of focusing (blue), at focusing (red), and upstream of focusing (green). Likewise, on the right, panels $(b,c,d)$ show the physical particle trajectories of these downstream, at focusing, and upstream particles respectively, normalized by $k_c$. Note that the total transport during focusing (red) is much greater than that away from focusing, contrary to linear theory \eqref{lineartransport} (dashed line) which states that all particles should experience the same transport.}
    \label{fig:trajectories}
\end{figure}

Figure (\ref{fig:trajectories}) shows a typical output of the surface particle trajectories during one such focusing event with bandwidth $\Delta = 0.8$ and linear prediction of maximum slope at focusing $S = 0.27$. For particles far downstream and upstream of focusing, represented by the blue and green curves respectively, their trajectories evolve gradually as the packet passes over. Their measured total transport $\delta x$, represented by the difference from their final and initial positions, mostly follows linear theory \eqref{lineartransport}. In contrast, for the particles at or near the focusing location, highlighted by the red curve, the transport occurs in one short burst as the focused packet passes over, well surpassing the predictions of linear theory and violating the supposed spatial invariance of the transport. For each particle, the total Lagrangian transport $\delta x$ is computed by taking the horizontal position averaged over the final two seconds of the simulation, roughly two central wave periods, and subtracting from it the particle's fixed initial position $x_0$. Both here and for the rest of this paper, we only show results for particles that began and ended at rest (i.e., they experienced the full packet passing)  so that total transport $\delta x$ is unambiguous.

\begin{figure}
    \centering
    \includegraphics[width=0.8\textwidth]{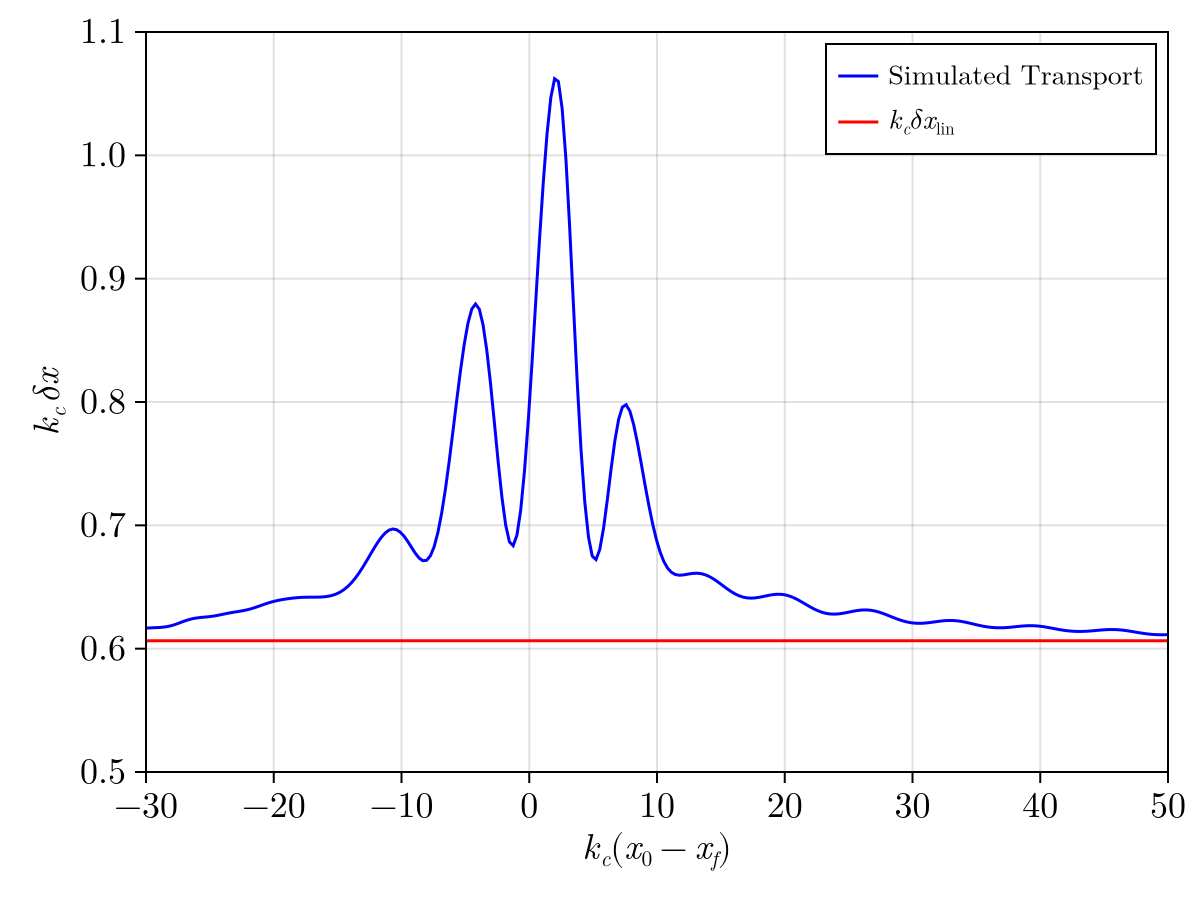}
    \caption{The total Lagrangian transport $\delta x$ of surface particles as a function of their initial distance from the linear prediction of maximum focusing $(x_0 - x_f)$, normalized by the central wavenumber $k_c$ for the same simulation as in figure (\ref{fig:trajectories}), $\Delta = 0.8$ and $S = 0.27$. The normalized linear prediction of the total transport $k_c \delta x_\text{lin}$ \eqref{lineartransport}, constant in space, is shown in red.}
    \label{fig:dxparticledep}
\end{figure}

Figure (\ref{fig:dxparticledep}) shows the computed total surface Lagrangian transport $\delta x$ as a function of its initial position relative to the linear focusing prediction, both normalized by the central wavenumber $k_c$ for the same simulation as in figure (\ref{fig:trajectories}). Plotted also is $\delta x_\text{lin}$, normalized by $k_c$, constant for each particle. From figure (\ref{fig:dxparticledep}), we see a strong spatial dependence of the total Lagrangian transport, with a maximum transport 75\% higher than linear theory predicts. At larger values of $\Delta$, where the physical packet width at focusing is smaller, the maximum transport was even found to be up to double that of linear theory. These are surprising results as it might be expected that any higher order corrections to linear theory would be necessarily small. Here we show that these corrections are comparable in magnitude to the linear prediction and exhibit a strong spatial dependence. In addition to the general increase around the focusing location, all simulations have oscillations in their transport curve near focusing with a spatial periodicity that matches the wavelength of the central wave. To compare these results with laboratory experiments, we also introduce a measure of the mean surface transport over the focusing region following \cite{sinnis21},
\begin{equation}
    \langle \delta x \rangle = \frac{1}{x_{02}-x_{01}}\int_{x_{01}}^{x_{02}} \delta x \, \dd x_{0} \, ,
\end{equation}
where in our study we define $x_{01}$ and $x_{02}$ as the first and last points respectively where the deviation of the transport from linear theory $(\delta x - \delta x_\text{lin})$ exceeds 10\% of its maximum value. In this case the mean transport is $23.6\%$ higher than linear theory.

While these simulations provide the first detailed account of the increased transport of steep non-breaking focusing wave packets, this study was motivated by earlier laboratory experiments \citep{lenain2019,sinnis21}. These wave tank experiments measured the spatially varying surface transport in primarily breaking focusing wave packets described by \eqref{focusEta}, with several steep non-breaking cases included for comparison. They found that wave breaking produces a large local increase to the surface transport. Wave breaking, in this case, breaks both the translational symmetry of the system and the transport in an obvious way. However, this symmetry breaking is also present for non-breaking focusing waves, allowing for a spatially dependent non-breaking transport which can be seen for example in \cite[][figure 5]{sinnis21}.

\begin{figure}
    \centering
    \includegraphics[width=0.85\textwidth]{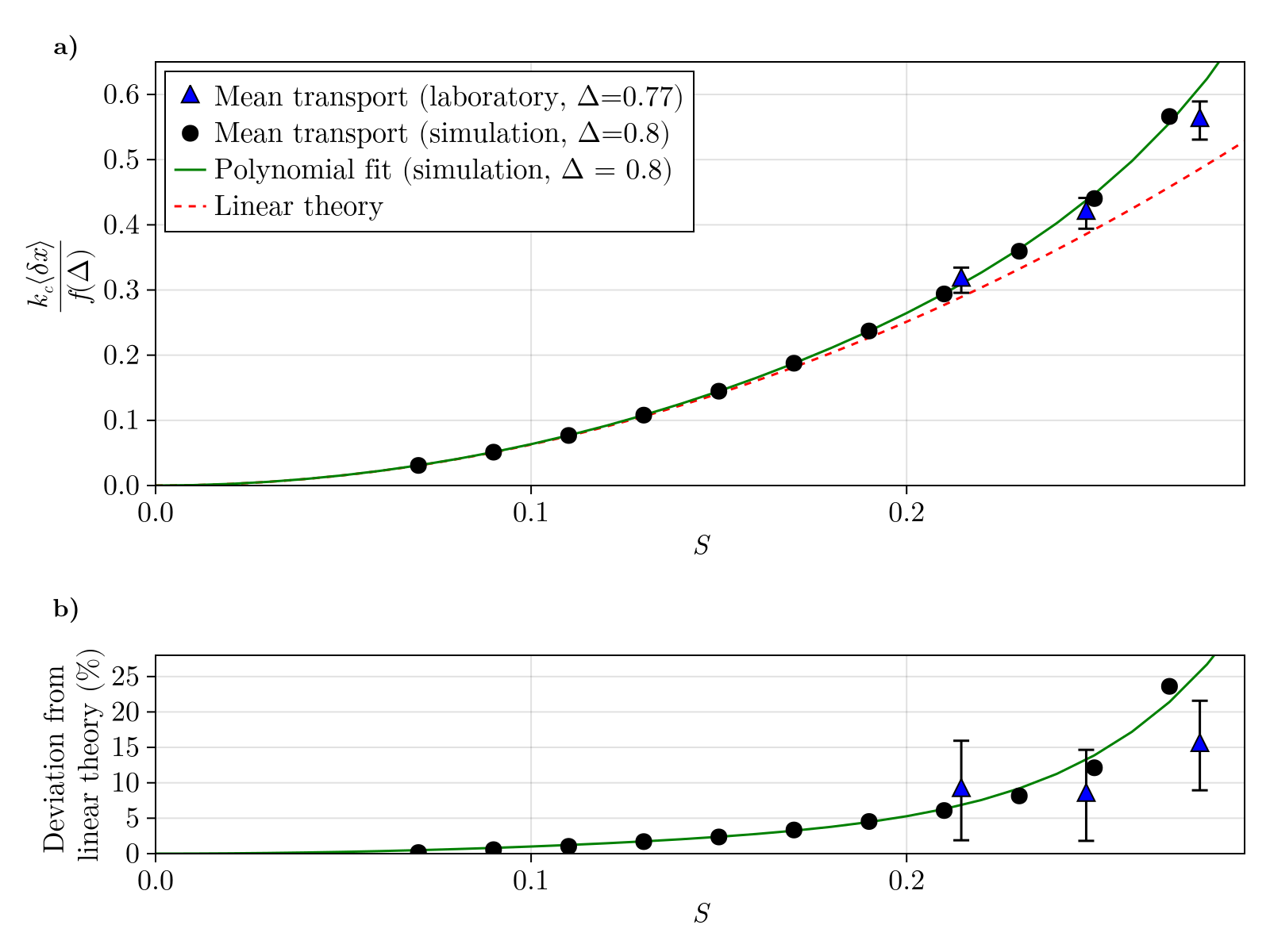}
    \caption{Mean surface transport $\langle \delta x \rangle$ as a function of the linear prediction of maximum wave slope $S$. Panel $$(a)$$ shows the mean transport normalized by the central wavenumber $k_c$ and linear bandwidth dependence $f(\Delta)$ so that the prediction of linear theory \eqref{lineartransport} (red) collapses to a single curve for both the simulation and laboratory parameters. A polynomial fit of the discrete simulation points is shown in green. Panel $(b)$ shows the same data plotted as a percentage increase from linear theory.}
    \label{fig:labdata}
\end{figure}

Figure (\ref{fig:labdata}) provides a direct comparison between the mean surface transport both observed in the laboratory and computed via simulation against the predictions of linear theory derived above plotted as a function of $S$. A particular bandwidth case $\Delta = 0.77$ for the laboratory data was chosen as it most closely approximates deep-water theory since the wave tank experiments were necessarily performed in a finite tank of mean water depth $\ell = 0.5 \, m$, which required using the full dispersion relationship $\omega^2 = g k \tanh(k \ell)$. Higher bandwidth packets contain longer wavelength waves which are modulated to a greater extent. For this particular bandwidth case, $\delta x_{\text{lin}}$ was $5\%$ less than its value in infinitely deep water. Plotted also is the mean transport computed from simulation for the nearest bandwidth case $\Delta = 0.8$. To compare the laboratory data and simulations, in \ref{fig:labdata}$(a)$ all data is normalized by the central wavenumber $k_c$ and linear bandwidth dependence $f(\Delta)$ so that the linear prediction is coincident for both cases. For the laboratory case, this requires numerically computing $f(\Delta)$ as the full dispersion relationship cannot be inverted in closed form. Additionally, a polynomial fit to the normalized simulated mean transport (green) is shown to guide the eye. From figure \ref{fig:labdata}$(a)$, it is clear that while linear theory accurately captures the transport at low slopes, there are significant increases to the mean Lagrangian transport when focusing wave packets become steep, validated by both experiment and simulation. To better visualize these enhancements, figure \ref{fig:labdata}(b) plots the same data instead as a percentage deviation from linear theory, where it can be seen that these mean enhancements are of comparable magnitude to the linear theory itself.

\begin{figure}
    \centering
    \includegraphics[width=0.7\textwidth]{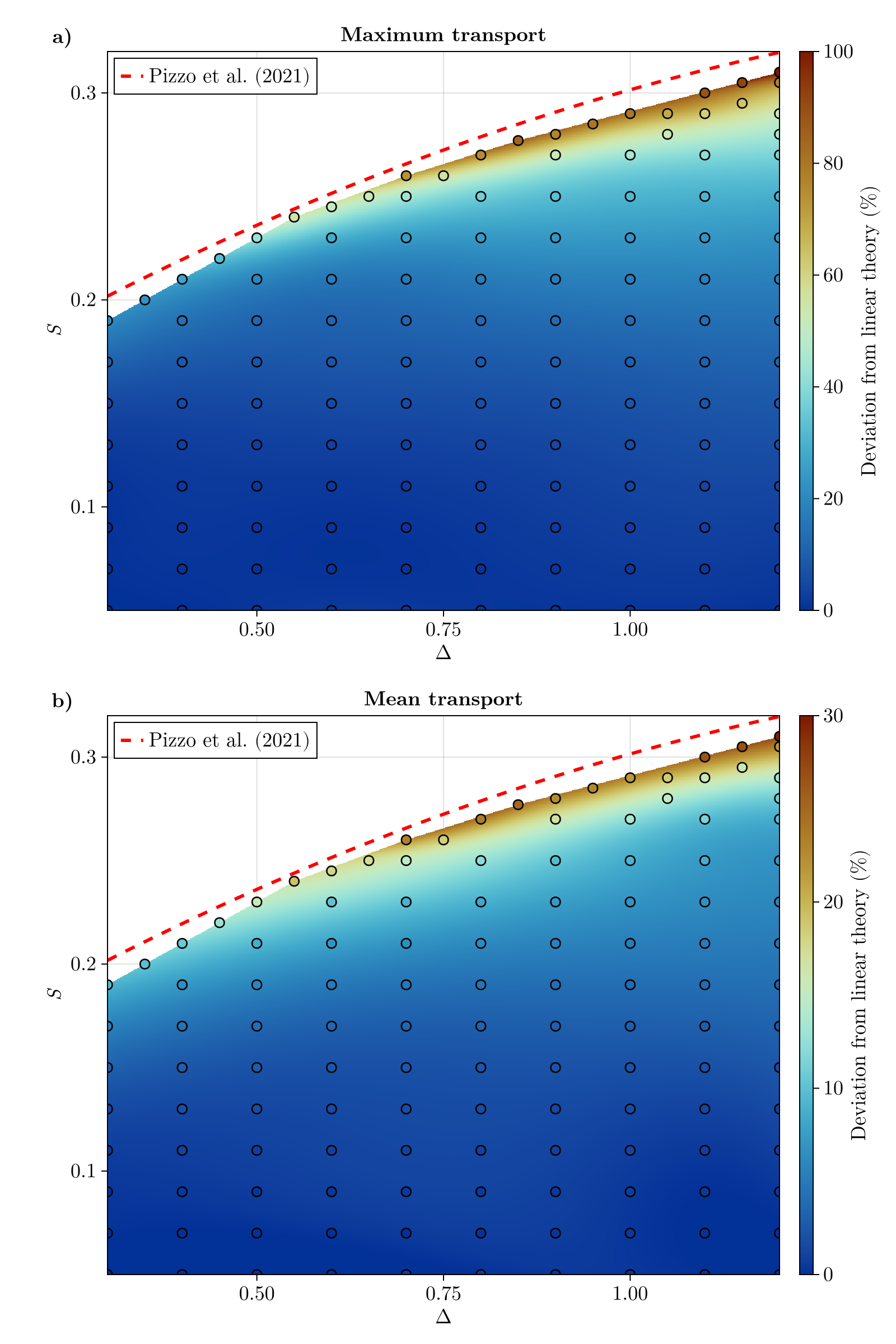}
    \caption{Percentage increases of the maximum $(a)$ and mean $(b)$ surface Lagrangian transport relative to linear theory \eqref{lineartransport} for numerically simulated focusing wave packets as a function of parameter space $(S,\Delta)$. Discrete simulation runs are shown via colored markers, with interpolated values in between. Note the two distinct colorbar scalings for panels $(a,b)$. The red line outlining the parameter space represents the breaking slope threshold numerically determined by \cite{pizzo2021b} which we found to be consistent with our simulations.}
    \label{fig:alldata}
\end{figure}

The enhancements of both the maximum and mean surface transport relative to linear theory for all simulations are shown in figure (\ref{fig:alldata}) as discrete points, with interpolated values in between. The dashed red line indicates the breaking slope threshold $S_*$ numerically determined by \cite{pizzo2021b} for equivalently defined deep-water packets. While linear theory accurately predicts the transport for low values of $S$, significant enhancements occur as waves steepen. Individual particles, shown in figure \ref{fig:alldata}$(a)$, can be transported up to twice as far as linear theory predicts, with the mean transport over the focusing region surpassing linear theory by up to $30\%$ as shown in figure \ref{fig:alldata}$(b)$. While the enhancements to the surface Lagrangian transport primarily scale with increasing $S$, there is also a noticeable $\Delta$ dependence close to the breaking threshold. To investigate why these spatially varying enhancements occur when waves steepen, we next turn to a theoretical derivation of the local mean Lagrangian drift for narrow-banded waves.

\section{Wave packets in the Lagrangian frame} \label{packets}

We begin by considering a unidirectional wave packet with a characteristic wavenumber $k_0$ and frequency $\omega_0=\sqrt{g k_0}$, and for simplicity normalize our units with new primed variables
\begin{equation}
    x' = k_0 x \, , \quad y' = k_0 y \, , \quad p' = \frac{\omega_0^2}{k_0^2} p \, , \quad \tau' = \omega_0 \tau \, , \quad \alpha' = k_0 \alpha \, , \quad \beta' = k_0 \beta \, ,
\end{equation}
which we henceforth drop for clarity of presentation. Using $\epsilon$ as a small steepness parameter, analogous to $S$, we start with nondimensionalized monochromatic waves
\begin{align}
    x &= \alpha + \frac{1}{2}\bigg(i\epsilon \F e^{i \theta_0} e^{\beta} + \complexconj \bigg) \, , \label{xpacket} \\
    y &= \beta + \frac{1}{2} \bigg(\epsilon \F e^{i \theta_0} e^{\beta} + \complexconj \bigg) \, , \label{ypacket}\\
    p &= -\beta \label{ppacket}\, ,
\end{align}
where $\theta_0 = \alpha - \tau$, $\F$ is an $O(1)$ nondimensional complex amplitude and $\complexconj$ indicates the complex conjugate. At this order $\F$ is constant and the solutions represent monochromatic plane waves identical to \eqref{monoX} -- \eqref{monoP}. To account for the effects of finite bandwidth, we allow for this complex amplitude to vary slowly in space and time, so that $\F = \F(\A,T)$ where we have introduced the new slow variables
\begin{equation}
    \A = \gamma  \alpha \, , \qquad T = \gamma \tau \, ,
\end{equation}
where $\gamma$ is another small parameter which is proportional to the normalized bandwidth $\Delta$. We separate $\gamma$ from $\epsilon$ to show how finite steepness and bandwidth individually affect the solutions following the approach of \cite{van_den_bremer_lagrangian_2016}, but assume both to be small parameters of the same asymptotic ordering so that a second order quantity, for example, describes terms proportional to any of the following: $\epsilon^2$, $\epsilon \gamma$, or $\gamma^2$. The general procedure for computing higher order solutions is to expand $x$, $y$ and $p$ in a standard asymptotic series,
\begin{align}
   x &= \alpha + \frac{1}{2}\bigg( i \epsilon \F e^{i \theta_0} e^{\beta} + \complexconj \bigg) + \sum_n \sum_m \epsilon^n \gamma^m x^{(n,m)}(\alpha,\beta,\tau,\A,T)\, , \label{xpacket2} \\
    y &= \beta + \frac{1}{2} \bigg(\epsilon \F e^{i \theta_0} e^{\beta} + \complexconj \bigg) + \sum_n \sum_m \epsilon^n \gamma^m y^{(n,m)}(\alpha,\beta,\tau,\A,T)\, , \label{zpacket2}\\
    p &= -\beta + \sum_n \sum_m \epsilon^n \gamma^m p^{(n,m)}(\alpha,\beta,\tau,\A,T)\label{ppacket2}\, .
\end{align}
Inserting \eqref{xpacket2} -- \eqref{ppacket2} into the Euler equations \eqref{xEuler}--\eqref{yEuler}, the irrotational condition \eqref{irrotational} and the continuity gauge choice $\J = 1$, and grouping terms by powers of $\epsilon$ and $\gamma$, we obtain a set of linear equations at second-order. Solving them along with the relevant boundary conditions yields
\begin{align}
    x &= \alpha + \frac{1}{2}\bigg(\bigg[i\epsilon \mathcal{F} + \epsilon \gamma \beta \mathcal{F}_{A} \bigg] e^{i \theta_0}e^\beta + \complexconj \bigg)
    + \epsilon^2 e^{2\beta}|\mathcal{F}|^2 \tau\, ,\label{packet2ndx} \\
    y &= \beta + \frac{1}{2}\bigg(\bigg[\epsilon \mathcal{F} -i \epsilon \gamma \beta \mathcal{F}_{A}\bigg] e^{i \theta_0}e^\beta + \complexconj \bigg) + \frac{\epsilon^2}{2} |\mathcal{F}|^2 e^{2 \beta} \, , \label{packet2ndy}\\
    p &= -\beta + \frac{1}{2}\bigg(\epsilon \gamma (i \mathcal{F}_A + 2i \mathcal{F}_T) e^{i \theta_0}e^\beta + \complexconj\bigg)  \, ,\label{packet2nd}
 \end{align}
where the oscillatory motion now no longer decays purely exponentially with depth, similar to what is found in the Eulerian frame for narrow-banded packets \citep{Yuen1975,Pizzo2016}. Just as with monochromatic waves, a second-order mean Lagrangian drift is required to enforce irrotational flow, although here its strength is set by the local squared magnitude of the packet envelope \citep{van_den_bremer_lagrangian_2016,haney_radiation_2017}. The presence of the waves also raises the potential energy of the fluid, as seen in \eqref{packet2ndy} through the Lagrangian mean water level. The condition that pressure vanishes at the sea surface requires $\F_T +\tfrac{1}{2}\F_\A = 0$ which simply means that to lowest order the envelope translates with the non-dimensional group velocity, which in deep water is half of the phase velocity.

As reported by \cite{Buldakov2006} for monochromatic waves, directly continuing this asymptotic expansion to third order yields nonphysical oscillatory terms in $x$ and $y$ which grow secularly in time. To obtain uniformly valid particle trajectories, \cite{Clamond2007} identified that the phase of the waves must be Doppler-shifted by the mean Lagrangian drift, effectively renormalizing the phase. This correction is necessary since two particles that are initially in phase (same $\alpha$, different $\beta$) gradually move out of phase at long times due to the vertically sheared mean Lagrangian drift transporting one more than the other. Since our system must reduce to monochromatic waves when the envelope is constant in space, we adopt the same renormalization so that, after shifting the phase of the carrier wave, so that we have
\begin{equation}
    \theta = \alpha - \tau + \epsilon^2 e^{2\beta}|\F|^2 \tau \, ,\label{thetadef}
\end{equation}
solutions valid to third order are
\begin{multline}
    x = \alpha + \frac{1}{2}\bigg(\bigg[i\epsilon \mathcal{F} + \epsilon \gamma \beta \mathcal{F}_{\A} -\epsilon\gamma^2  i \frac{\beta^2}{2}\mathcal{F}_{\A \A} + \epsilon^3 2i |\mathcal{F}|^2 \mathcal{F}e^{2\beta}\bigg] e^{i \theta}e^\beta + \complexconj \bigg) \\
    + \int \langle x_\tau \rangle\,  \dd \tau\, , \label{x3}
\end{multline}
\begin{multline}
    y = \beta + \frac{1}{2}\bigg(\bigg[\epsilon \mathcal{F} -i \epsilon \gamma  \beta \mathcal{F}_{\A}  -\epsilon \gamma^2 \frac{\beta^2}{2}\mathcal{F}_{\A \A} +  \epsilon^3|\mathcal{F}|^2 \mathcal{F}e^{2\beta}\bigg] e^{i \theta}e^\beta + \complexconj \bigg)
    + \frac{\epsilon^2}{2} |\mathcal{F}|^2 e^{2 \beta} \\+ \epsilon^2 \gamma \bigg(-\frac{i}{2}\bigg(\beta + \frac{1}{2}\bigg)(\F_\A \F^* - \F \F^*_\A) \bigg) e^{2\beta} + \int \langle y_\tau \rangle \, \dd \tau \, ,\label{y3}
\end{multline}
\begin{align}
    p = -\beta + \frac{1}{2}\bigg(\bigg(\epsilon \gamma (i \mathcal{F}_A + 2i \mathcal{F}_T) - \epsilon^3|\mathcal{F}|^2\mathcal{F} e^{2\beta} - \frac{1}{4}\epsilon \gamma^2 \mathcal{F}_{\A\A}\bigg)\bigg)e^{i \theta}e^\beta + \complexconj \bigg) \, ,\label{p3}
 \end{align}
where the mean Lagrangian drift, left in general form here, is formally derived in the following subsection. It is important to note that only the second-order mean terms are required to fully constrain the third-order orbital motion. From the condition of vanishing pressure at the free surface we derive an equation governing the evolution of the wave envelope,
\begin{equation} \label{temporalNLSE}
    i\bigg(\F_T + \frac{1}{2}\F_\A\bigg) -  \Bigg(\gamma\frac{1}{8}\F_{\A\A} + \frac{\epsilon^2}{\gamma}\frac{1}{2}|\F|^2 \F\Bigg)  = 0\, ,
\end{equation}
which reduces to the classical nonlinear Schrödinger equation (NLSE) for narrow-banded irrotational waves when $\epsilon = \gamma$ \citep{Zakharov1968}. This equation \eqref{temporalNLSE} has a number of conserved quantities in time. These are, as they are commonly referred to in the literature, the linear wave energy
\begin{equation}\label{NLSEenergy}
    \mathcal{E} = \int |\F|^2 \, \dd \A \, ,
\end{equation}
the mean wavenumber
\begin{equation}\label{NLSEmomentum}
    \mathcal{P} = \int i(\F^*\F_\A - \F \F_\A^*) \, \dd \A \, ,
\end{equation}
and the Hamiltonian
\begin{equation}\label{NLSEhamiltonian}
    H = \int \frac{1}{4}\epsilon^2|\F|^4 - \frac{1}{8}\gamma^2|\F_\A|^2 \, \dd \A \, ,
\end{equation} 
which arise via Noether's theorem from symmetries of the NLSE action to phase shifts, spatial translation, and time translation respectively \citep{sulem1999nonlinear}. 

\subsection{The mean Lagrangian drift of steep narrow-banded waves}

Because we enforced a scale separation between the fast orbital motion and the slow envelope evolution through the small parameter $\gamma$, the averaging operator $\langle \cdot \rangle$ is defined as a spatial convolution over an intermediate scale -- large enough to remove the fast oscillations but small enough to retain slow envelope modulations. Spatial convolutions commute with the curl, so that from \eqref{generalDrift} we know that the curl of the mean Lagrangian drift is exactly equal to the curl of the mean pseudomomentum. The pseudomomentum itself is computed from products of the Lagrangian particle displacements $(\xi,\zeta)$ whose leading-order contributions are $O(\epsilon)$; consequently, the fourth-order pseudomomentum is determined entirely by third-order terms. It is easy to check that the second-order mean terms in \eqref{packet2ndx}--\eqref{packet2ndy} only contribute to the curl of the mean pseudomomentum beginning at fifth order, so that only the oscillatory terms are required. Direct evaluation of this quantity from \eqref{x3}--\eqref{y3} shows that the mean Lagrangian drift must satisfy
\begin{multline}
    \frac{\d \langle y_\tau \rangle}{\d \alpha} - \frac{\d \langle x_\tau \rangle}{\d \beta}= -2 \epsilon^2 |\F|^2 e^{2\beta} + \epsilon^2\gamma \bigg(2i (1 + \beta) \big(\F_\A \F^* - \F \F_\A^*\big) - i \big(\F_T\F^* - \F \F_T^*\big)\bigg)e^{2\beta} \\
    - 8 \epsilon^4 |\F|^4 e^{4\beta} + \epsilon^2 \gamma^2 \bigg(\big(\tfrac{1}{2} + \tfrac{5}{2}\beta + \beta^2\big)|\F|^2_{\A\A} e^{2\beta} - \big(4 + 10\beta + 4\beta^2\big)|\F_\A|^2 e^{2\beta}\bigg) \, \, ,\label{curlpseudo}
\end{multline}
where we have used the identity
\begin{equation}
    \big(\F_{\A\A}\F^* + \F \F_{\A\A}^*\big)= |\F|^2_{\A\A} - 2|\F_\A|^2 \, ,
\end{equation}
to consolidate various terms. In its current form \eqref{curlpseudo} is underdetermined since we cannot ignore $\d_\alpha\langle y_\tau \rangle$ at higher orders. Although the fluid is incompressible, the Lagrangian velocity need not be divergence-free \eqref{incompressibility}. As pointed out by \cite{vanneste_stokes_2022}, the fact that waves modify the potential energy of a fluid implies a changing center of mass, which, when paired with the bottom boundary condition, requires a divergent Lagrangian mean flow. We can account for this, however, in the Lagrangian mean water level, whose value can be computed from the Jacobian \eqref{Jacobian} independently of the mean Lagrangian drift to fourth-order from \eqref{x3}--\eqref{y3},
\begin{multline} \label{MWL}
    \langle y\rangle_{\text{MWL}} = \frac{1}{2}\epsilon^2|\F|^2 e^{2\beta} - \frac{1}{2}\epsilon^2 \gamma i \bigg(\beta + \frac{1}{2}\bigg)\big(\F_{\A}\F^* - \F \F_\A^*\big) e^{2\beta} +  \frac{3}{2}\epsilon^4 |\F|^4 e^{4\beta}\\ + \epsilon^2 \gamma^2 \bigg( (\beta + \beta^2)|\F_\A|^2 + \frac{1}{8}(1 - 2\beta -2\beta^2 ) |\F|_{\A\A}^2 \bigg)e^{2\beta} \, .
\end{multline}
The Lagrangian mean water level \eqref{MWL}, or more aptly changes thereof, fully constrain the divergent part of the total mean Lagrangian velocity. This can be seen by taking the average of the incompressibility condition \eqref{incompressibility}, which from \eqref{x3}, \eqref{y3} and \eqref{MWL} imply
\begin{equation}
 \langle x_\tau\rangle_\alpha + \langle y_\tau \rangle_\beta =  O(\epsilon^4 \gamma)\, ,\label{incompressible}
\end{equation}
so that to fourth-order, the mean Lagrangian drift is divergence-free (at higher orders products of the mean Lagrangian drift enter \eqref{incompressible}). This allows us to define a streamfunction for the mean Lagrangian drift $\psi $ such that
\begin{equation}\label{meanstreamfunction}
    \langle x_\tau \rangle= -\psi_\beta \, , \qquad \langle y_\tau \rangle = \psi_\alpha \, .
\end{equation}
The above equation \eqref{curlpseudo} does not distinguish between the mean Lagrangian drift or the Lagrangian mean water level, so to only constrain the drift we must account for the mean water level's contribution to the curl, which only emerges at fourth-order,
\begin{equation}
    \d_\alpha \d_\tau (\langle y \rangle_\text{MWL}) = \frac{1}{2}\epsilon^2 \gamma^2|\F|^2_{\A T} \approx -\frac{1}{4}\epsilon^2 \gamma^2 |\F|^2_{\A\A} \, ,
\end{equation}
where for simplicity we have used the fact that to lowest order, $\F_T = -\tfrac{1}{2}\F_\A$. Combining \eqref{curlpseudo} with \eqref{meanstreamfunction} therefore results in
\begin{multline}
    \nabla^2_{\bm{\alpha}} \psi = -2 \epsilon^2 |\F|^2 e^{2\beta} + \epsilon^2\gamma \bigg(2i (1 + \beta) \big(\F_\A \F^* - \F \F_\A^*\big) - i \big(\F_T\F^* - \F \F_T^*\big)\bigg)e^{2\beta} \\
    - 8 \epsilon^4 |\F|^4 e^{4\beta} + \epsilon^2 \gamma^2 \bigg(\big(\tfrac{3}{4} + \tfrac{5}{2}\beta + \beta^2\big)|\F|^2_{\A\A} e^{2\beta} - \big(4 + 10\beta + 4\beta^2\big)|\F_\A|^2 e^{2\beta}\bigg) \, , \label{meanflowvorticity}
\end{multline}
which is just the linear Poisson equation. One can interpret \eqref{meanflowvorticity} as equating the vorticity of the mean Lagrangian drift to the curl of the mean pseudomomentum, which acts here as a wave-induced source of vorticity \citep{Salmon2020}. It is important to reiterate that despite this interpretation, the vorticity of the fluid is still exactly zero everywhere in the fluid. The monochromatic mean Lagrangian drift shows how a mean flow that is sheared in the Lagrangian reference frame can still describe perfectly irrotational flow.

To close the system, we require boundary conditions on $\psi$. The first one is simple: the flow vanishing at infinite depth implies $\psi \rightarrow 0$ as $\beta \rightarrow -\infty$. The surface boundary condition is more subtle, as it was implicitly introduced in $\S$\ref{lintheory} from the fact that the surface of the fluid is always defined by particles with $\beta=0$. This is the Lagrangian equivalent of the kinematic boundary condition, \eqref{KBC} in the Eulerian frame, which in plain language states that particles which start at the surface always remain at the surface. From the perspective of the mean Lagrangian drift, there can therefore be no mass flux through the surface, making it a streamline which we can without loss of generality set to $\psi = 0$ at $\beta=0$. This is not to say that there can be no mean vertical motion, only that any such motion at the surface must correspond with changes to the surface geometry, which is already set by the Lagrangian mean water level. Slow changes in the Lagrangian mean water level, found by taking a time derivative of \eqref{MWL}, can be interpreted as a `vertical drift' which several studies investigate \citep[e.g.,][]{vanneste_stokes_2022}. We choose to separate these effects due to their distinct dynamical origins; the mean Lagrangian drift is fundamentally set by the vorticity (or lack thereof), whereas the Lagrangian mean water level is constrained by the geometry of material curves and would be present even in the absence of the mean Lagrangian drift, such as in the rotational \cite{Gerstner1802} wave. 

The full solution to \eqref{meanflowvorticity} can be found in appendix \ref{appendixA}, though its general character is determined solely by considering the pseudomomentum forcing term. To leading order, this forcing is negative and concentrated near the surface. Assuming the wave packet has a finite width, so that the streamlines must be closed, a clockwise circulation will develop which moves with the packet. This circulation presents itself as a strong jet near the surface, as determined in \eqref{packet2ndx} as the classical mean Lagrangian drift, but also includes a slow deep return flow in a direction opposite to that of wave propagation that is well known in the literature \citep{LH1962,Mcintyre1981,Salmon2020,pizzowagner2025}. These studies, however, only constrain the lowest order mean flow response. 

At higher orders, additional forcing terms arise that modify the structure of the mean Lagrangian drift. The third-order forcing terms in \eqref{meanflowvorticity} depend on the quantities $i(\F_\A\F^* - \F \F_\A^*)$ and $i(\F\F_T^* - \F \F_T^*)$ which represent local fluctuations to the wavenumber and frequency from their characteristic values $k_0$ and $\omega_0$ respectively. These terms therefore account for modifications to the mean Lagrangian drift associated with local variations in the phase speed. 

At fourth-order, there is a forcing term proportional to the fourth power of the envelope magnitude, which will always act to strengthen the near-surface mean Lagrangian drift, particularly during focusing when its magnitude is most pronounced. The quantity $|\F|^2_{\A\A}$ is the curvature of the squared envelope magnitude; it enhances the forcing where the curvature of the envelope is most negative (near the packet center), and reduces it at the edges where the curvature is positive. Owing to the nontrivial vertical dependence, this effect is reversed at depth. The final contribution, proportional to $|\F_\A|^2$, similarly enhances the forcing near the surface. Together, these fourth-order corrections provide enhancements to the forcing of the mean Lagrangian drift near the surface in regions where wave envelopes are both steep and concave, precisely the conditions present during focusing which led to the greatest observed transport. 

Despite the complexity of the full solution, the mean Lagrangian drift at the surface can be written explicitly as (see appendix \ref{appendixA}),
\begin{multline}
    \langle x_\tau \rangle\Big|_{\beta=0} =\epsilon^2 |\F|^2 + \epsilon^2\gamma \bigg(\frac{1}{2}i \big(\F_\A \F^* - \F \F_\A^*\big) - \frac{1}{2} i\big(\F_T\F^* - \F \F^*_T\big) +\frac{1}{2}\mathcal{H}\big(|\F|^2_\A\big)\bigg) \\
    + 2\epsilon^4 |\F|^4 + \epsilon^2 \gamma^2 \bigg( \frac{1}{2}|\F_\A|^2 - \frac{1}{4}|\F|^2_{\A\A} + \frac{1}{4}i \d_\A\mathcal{H}\big(\F_{T}\F^* - \F\F_{T}^*\big)\Big)\bigg) \, , \label{driftsurface}
\end{multline}
where $\mathcal{H}$ represents the spatial Hilbert transform. 

To check that our solutions reduce to known results for steep monochromatic waves, we see what happens when the envelope $\F$ is constant in space. From \eqref{temporalNLSE}, we have
\begin{equation}
    i \F_T = \frac{\epsilon^2}{\gamma}\frac{1}{2}|\F|^2 \F \, ,
\end{equation}
which admits the exact solution (recalling $T = \gamma \tau$)
\begin{equation} \label{stokespacketsolution}
    \F = \F_0 e^{-\frac{1}{2}i \epsilon^2 |\F_0|^2 \tau} \, ,
\end{equation}
where $\F_0$ is a complex constant, still $O(1)$. This slow time modulation to $\F$ is just the classical Stokes correction to the phase speed for finite amplitude waves \citep{Stokes1847}. Assuming without loss of generality that $\F_0 = 1$ (the physical dimensions can be added later), inserting \eqref{stokespacketsolution} into the trajectories \eqref{x3}--\eqref{p3} yields
\begin{align}
    x &= \alpha - (\epsilon + 2\epsilon^3 e^{2\beta})e^\beta\sin(\alpha - (c-U(\beta))\tau) + U(\beta)\tau , \\
    y &= \beta + (\epsilon + \epsilon^3 e^{2\beta}) e^\beta \cos(\alpha - (c-U(\beta))\tau) + \frac{1}{2}\epsilon^2 e^{2\beta} \, ,\label{ymono} \\ 
    p &= -\beta + \epsilon^3(e^\beta - e^{3\beta})\cos(\alpha - (c-U(\beta))\tau) \, ,
\end{align}
where $c = \Big( 1 + \tfrac{1}{2} \epsilon^2 \Big)$ is the nonlinear, nondimensionalized phase speed, and $U(\beta)$ is the mean Lagrangian drift, governed by
\begin{equation} \label{monostreamfunction}
    \nabla_{\bm{\alpha}}^2 \psi = -2\epsilon^2 e^{2\beta} - 8 \epsilon^4 e^{4\beta} - \epsilon^4 e^{2\beta} \, ,
\end{equation}
which only depends on $\beta$. Therefore, \eqref{monostreamfunction} reduces to an ordinary differential equation, and by simple integration the solution becomes
\begin{equation}\label{monodriftClamond}
    U(\beta) = -\psi_\beta = \epsilon^2 \underbrace{\Big(1 + \tfrac{1}{2}\epsilon^2\Big)}_ce^{2\beta} + 2\epsilon^4 e^{4\beta}
\end{equation}
While these are uniformly valid solutions that match those of \cite{Clamond2007}, he defines his small steepness parameter $\epsilon'$ to be equal to $k_0 h/2$, where $h$ is the crest to trough distance at the surface. Reading the surface crest to trough distance from our solution \eqref{ymono} implies the relationship between these small parameters is
\begin{equation}
    \epsilon' = \epsilon + \epsilon^3\, ,
\end{equation}
so that using the standard definition of wave steepness $\epsilon'$, the drift at the surface can be expressed as
\begin{equation}
    U(0) = \epsilon'^2 e^{2\beta} c' \, \,, \qquad c' = \Big(1 + \tfrac{1}{2}\epsilon'^2\Big) \, ,
\end{equation}
to fourth order in $\epsilon'$ which matches the results in the literature \citep{LH1987}. Because unsteady narrow-banded waves do not have an unambiguous geometric reference such as crest to trough height, we must instead settle for its more basic definition above based on the steepness of the first order Lagrangian expansions.

\subsection{Comparison with simulation}

As a test of the above theory, we directly apply our above expression for the local mean Lagrangian drift of steep, narrow-banded waves \eqref{driftsurface} to estimate the surface transport of focusing wave packets from measurements of the wave envelope, comparing these predictions to the results from our fully nonlinear simulations. At the surface the transport is given by integrating \eqref{driftsurface} in time,
\begin{multline} \label{theorytransport}
    \delta x = \int \epsilon^2 |\F|^2 + \epsilon^2\gamma \bigg(\frac{1}{2}i \big(\F_\A \F^* - \F \F_\A^*\big) - \frac{1}{2} i\big(\F_T\F^* - \F \F^*_T\big) +\frac{1}{2}\mathcal{H}\big(|\F|^2_\A\big)\bigg) \\
    + 2\epsilon^4 |\F|^4 + \epsilon^2 \gamma^2 \bigg( \frac{1}{2}|\F_\A|^2 - \frac{1}{4}|\F|^2_{\A\A} + \frac{1}{4}i \d_\A\mathcal{H}\big(\F_{T}\F^* - \F\F_{T}^*\big)\bigg) \, \dd \tau\, .
\end{multline}
To compare this prediction with the simulations, $\F$ is estimated by taking the spatial Hilbert transform of the vertical Lagrangian positions at each time. While this correctly estimates the magnitude of the envelope $\F$, its phase must be corrected by removing the phase of the carrier wave, which is given by \eqref{thetadef}. Because the theory outlined above is non-dimensional, all quantities must be dimensionalized by a characteristic wavenumber $k_0$ and frequency $\omega_0 = \sqrt{g k_0}$. Note that $k_0$ and $\omega_0$ need not be equivalent to the initially specified $k_c$ and $\omega_c$. Because the phase of $\F$ represents narrow-banded deviations from the phase of the carrier wave, we choose $k_0$ such that these deviations have zero mean when integrated in time. Once the envelope $\F$ is computed, all that remains is to compute the surface mean Lagrangian transport \eqref{theorytransport} (where $\epsilon$ and $\gamma$ are implicitly included in measurements of $\F$ and its slow derivatives) for each particle and integrate in time over the duration of the packet passing to estimate the total transport predicted by this theory. From this, we can compute the maximum and mean transport predicted by narrow-banded theory to compare directly with the simulations.

\begin{figure}
    \centering
    \includegraphics[width=1.0\linewidth]{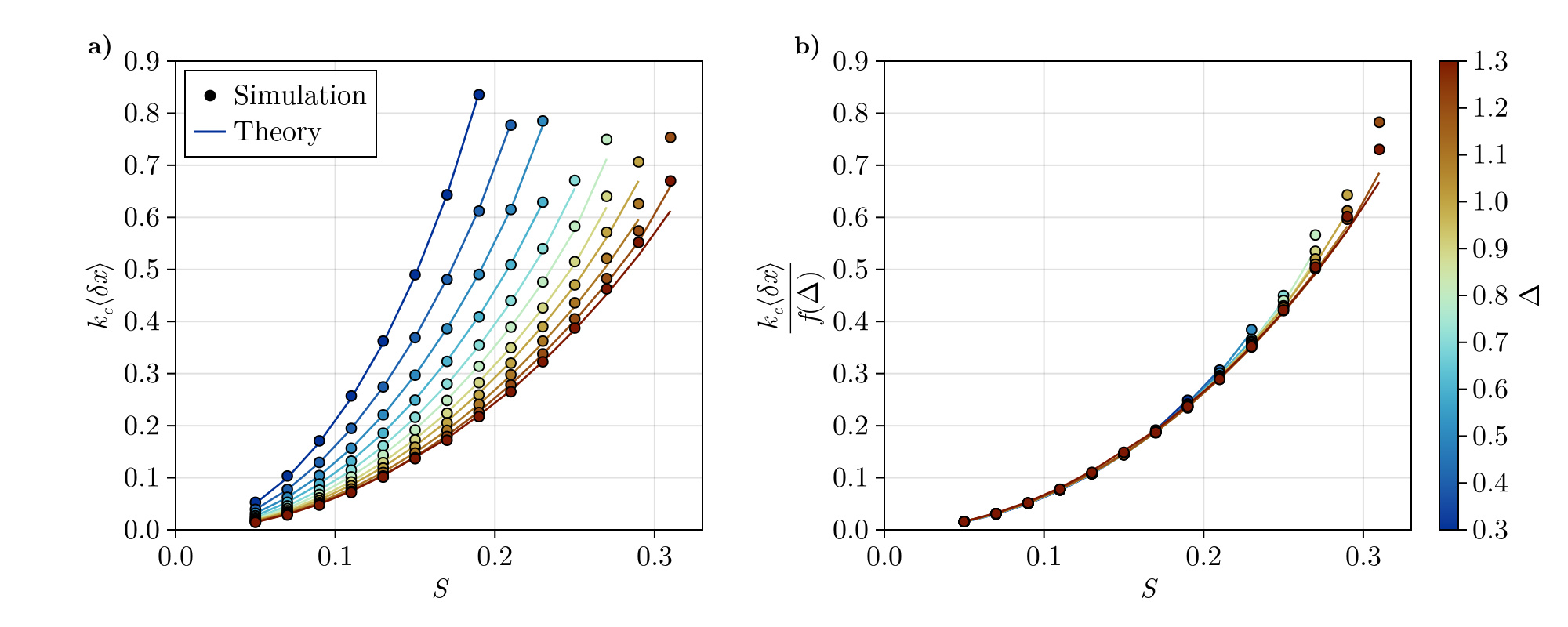}
    \caption{The mean surface transport $\langle \delta x \rangle$ within the focusing region as a function of $S$ computed both directly from simulation (circles) and using our higher-order theory \eqref{theorytransport} (lines) for each simulation. In $(a)$, $\langle \delta x \rangle$ is normalized by the central wavenumber $k_c$, and each line represents the theoretical prediction of mean transport for each bandwidth value. In $(b)$, $\langle \delta x \rangle$ is also normalized by the linear bandwidth dependence $f(\Delta)$ \eqref{transportapprox} which collapses the results. The theory performs best at lower values of $\Delta$ where the narrow-banded envelope assumption is most valid.}
    \label{fig:theoryprediction}
\end{figure}

Figure (\ref{fig:theoryprediction}) presents the mean surface transport obtained directly from the simulation, together with its theoretical prediction from \eqref{theorytransport} evaluated for each case. In figure \ref{fig:theoryprediction}(a), the mean transport is scaled only by the central wavenumber $k_c$, and good agreement between simulation and theory is found, especially for lower values of $\Delta$ and $S$. This is expected since \eqref{theorytransport} is valid to fourth order in the small parameters $(\epsilon,\gamma)$ which are proportional to $(S,\Delta)$ respectively. The theory still performs well across a wide range of parameter space, indicating that the narrow-banded wave approximation is able to capture the local enhancements to the mean Lagrangian drift. Figure \ref{fig:theoryprediction}(b) collapses these results, normalizing by both $k_c$ and the linear bandwidth dependence $f(\Delta)$, showing that the bandwidth dependence of the surface transport, even in steep packets, is still reasonably approximated by linear theory.

\section{Discussion} \label{discussion}

In this paper we investigated the mean Lagrangian drift for irrotational steep focusing surface gravity waves. By working directly in the Lagrangian reference frame, we derived a novel exact technique for constraining the mean Lagrangian drift in general wavy flows, illustrating its role as a spatially varying dynamic mean flow instead of as a passive byproduct of waves. Through a combination of numerical simulations and archived laboratory data, we showed that the surface Lagrangian transport in steep focusing waves can vary spatially and is significantly increased in regions of wave focusing. By performing a separation of scales analysis in the Lagrangian reference frame, we derived Lagrangian particle trajectories in narrow-banded steep wave fields, and derived a higher-order expression for the local mean Lagrangian drift and corresponding deep recirculation flow. The form of this expression suggests that wave focusing locally increases the surface drift. Comparing the predictions of this theory with the simulated results show that it captures a large portion of the observed enhancements especially at smaller bandwidths where the narrow-banded assumption holds. 

This study in general advocates for a more local interpretation of the mean Lagrangian drift. For irrotational flow, we showed that the curl of the mean Lagrangian drift is exactly equal to the curl of the mean pseudomomentum, which itself originates from correlations of wavy particle displacements. Consequently, spatial modifications to the wave orbital motion generate local variations in the mean pseudomomentum, which in turn drive corresponding variations in the mean Lagrangian drift. As evidenced by the particle trajectories within focusing wave packets, this has a profound affect on the way particles are transported -- those directly in the focusing region are transported in a rapid burst, whereas particles further downstream drift more slowly as the packet disperses. Modeling the mean Lagrangian drift as a dynamic mean flow which can vary in both space and time may better explain various processes such the enhanced horizontal diffusion due to waves \citep{herterich_horizontal_1982} and could provide more insight into how this mean flow interacts with the vorticity field to generate Langmuir circulation.

These results show that it is the local steepness of the wave field, not just the steepness of individual wave components, which sets the magnitude of these enhancements. That is, even if individual waves comprising a wave field are otherwise well described by linear theory, linear dispersion will consistently create localized focusing events that produce bursts of an increased near surface mean Lagrangian drift. While the likelihood of all wave components constructively interfering such as in the packets studied above may be low, the local steepness need only approach moderate values to begin to see these enhancements, which can occur so long as only some waves constructively interfere. Such focusing events should be commonplace in moderately developed seas, suggesting that both the magnitude and vertical dependence of the mean Lagrangian drift may be incorrectly estimated with models that ignore these effects.

\small
\vspace{3mm}

\noindent 
\textbf{Acknowledgements}
{We thank R. Salmon for many insightful conversations.}
\vspace{3mm}\\
\textbf{Funding}
{Blaser, Pizzo and Lenain were partially supported by NSF OCE-2219752, 2342714 and 2342716, NASA 80NSSC19K1037 (S-MODE) and 80NSSC23K0985 (OVWST) awards.}
\vspace{3mm}\\
\textbf{Competing Interests.}
{The authors report no conflict of interest.}
\vspace{3mm}\\
\noindent \textbf{Author Contributions.}\\
\textbf{Blaser} Formal analysis (lead); writing - original draft presentation (lead); writing - review and editing (lead); conceptualization (supporting). \textbf{Lenain} Formal analysis (supporting), review (supporting), conceptualization (supporting), Funding acquisition (lead). 
\textbf{Pizzo} Formal analysis (supporting), review (supporting), conceptualization (lead), Funding acquisition (lead).
\\

\noindent \textbf{Author ORCIDs.}\\
\orcidlink{0000-0002-3104-4589} Aidan Blaser \href{https://orcid.org/0000-0002-3104-4589}{https://orcid.org/0000-0002-3104-4589} \\
\orcidlink{0000-0001-9808-1563} Luc Lenain \href{https://orcid.org/0000-0001-9808-1563}{https://orcid.org/0000-0001-9808-1563}\\
\orcidlink{0000-0001-9570-4200} Nick Pizzo \href{https://orcid.org/0000-0001-9570-4200}{https://orcid.org/0000-0001-9570-4200}

\normalsize
\appendix

\section{Streamfunction for the mean Lagrangian drift}\label{appendixA}

Here we solve Poisson's equation for the streamfunction of the mean Lagrangian drift \eqref{meanflowvorticity} and compute its value at the surface. In $\S$\ref{packets}, we showed that after averaging over the fast orbital motion, the mean Lagrangian drift is divergence-free in label space to fourth-order, allowing for the introduction of a streamfunction for the mean flow with the following sign convention
\begin{equation}
    \langle x_\tau\rangle = -\psi_\beta \, , \qquad \langle y_\tau \rangle = \psi_\alpha \, .
\end{equation}
Using the exact equivalence between the the curl of the mean Lagrangian drift and the curl of the mean pseudomomentum \eqref{generalDrift}, this streamfunction was found to obey
\begin{multline}
    \nabla^2_{\bm{\alpha}} \psi = -2 \epsilon^2 |\F|^2 e^{2\beta} + \epsilon^2\gamma \bigg(2i (1 + \beta) \big(\F_\A \F^* - \F \F_\A^*\big) - i \big(\F_T\F^* - \F \F_T^*\big)\bigg)e^{2\beta} \\
    - 8 \epsilon^4 |\F|^4 e^{4\beta} + \epsilon^2 \gamma^2 \bigg(\big(\tfrac{3}{4} + \tfrac{5}{2}\beta + \beta^2\big)|\F|^2_{\A\A} e^{2\beta} - \big(4 + 10\beta + 4\beta^2\big)|\F_\A|^2 e^{2\beta}\bigg) \, , \label{poisson}
\end{multline}
within the fluid domain, which in label space is just the lower half plane. The system is closed with boundary conditions $\psi = 0$ at the surface $(\beta=0)$ and at infinite depth $(\beta \rightarrow -\infty)$. The above equation \eqref{poisson} is linear, meaning that we can separate $\psi$ into a homogeneous and particular solution,
\begin{equation}
    \psi = \psi_H + \psi_P \, . \label{psidecomp}
\end{equation}
The right-hand side of \eqref{poisson} only varies slowly in $\alpha$, so by inspection, the particular solution valid to fourth-order is
\begin{multline}
    \psi_P = -\frac{1}{2}\epsilon^2 |\F|^2 e^{2 \beta}+ \epsilon^2\gamma \bigg(-\frac{1}{4}i \big(\F_T\F^* - \F \F_T^*\big) + \frac{1}{2}i \beta \big(\F_\A\F^* - \F \F_\A^*\big)\bigg)e^{2\beta} \\
    - \frac{1}{2}\epsilon^4 |\F|^4 e^{4\beta} + \epsilon^2\gamma^2 \bigg(\frac{1}{16}\big(1 + 2\beta + 4\beta^2\big) |\F|^2_{\A\A} - \frac{1}{2}\big(\beta + 2\beta^2\big)|\F_\A|^2\bigg)e^{2\beta} \, .\label{particularsolution}
\end{multline}
From \eqref{psidecomp}, we see that $\psi_H$ must therefore obey
\begin{equation}
    \nabla^2_{\bm{\alpha}}\psi_H = 0 \, , \qquad \psi_H \Big|_{\beta=0} = -\Xi(\A,T) \, ,\label{homogeneousEq}
\end{equation}
where for brevity $\Xi(\A,T)$ is defined as the right-hand side of \eqref{particularsolution} evaluated at the surface. Since the homogeneous system \eqref{homogeneousEq} only depends on the slow variables $\A$ and $T$, we introduce a slow vertical variable $\mathcal{B}=\gamma \beta$ to make explicit that $\psi_H$ varies slowly with depth. The homogeneous solution still obeys the Laplace equation in these slow variables (i.e., $\d_{\A}^2 \psi_H + \d_{\mathcal{B}}^2\psi_H = 0)$, and because the domain is the lower half plane, we can use the theory of Poisson kernels to immediately write the full solution
\begin{equation}
    \psi_H = \frac{1}{\pi}\int_{-\infty}^\infty \frac{\Xi(\A',T)\mathcal{B}}{(\A'-\A)^2 + \mathcal{B}^2} \, \dd \A' \, . \label{homogeneoussolution}
\end{equation}
Looking at the form of these solutions, it is clear that $\psi_P$ describes the majority of the mean Lagrangian drift near the surface, manifesting as a jet beneath the wave envelope as expected. However, the streamlines of $\psi_P$ intersect the surface for a compact packet, which violates the kinematic boundary condition that surface particles remain at the surface. This is remedied by $\psi_H$, which enforces the boundary condition and describes a deep recirculation flow opposite in direction to packet propagation, generating downwelling and upwelling beneath the front and back of the packet respectively. Combined, the streamlines of $\psi$ are all closed beneath the surface, indicative of one half of a Bretherton dipole flow \citep{bretherton_mean_1969}. Finally, we note that in a box which contains the entire wave packet, such that $\psi \rightarrow 0$ on all boundaries, it is easy to show through the divergence theorem that even for this higher order solution the total integrated momentum vanishes, which is a well established result in the literature \citep{Mcintyre1981,pizzowagner2025}.

From this solution, we can directly compute the mean Lagrangian drift at the surface. For $\psi_P$, it is simple enough to take a derivative with respect to $\beta$. For $\psi_H$, we can use the fact that it obeys Laplace's equation to say it is the imaginary part of an analytic complex potential $\chi = \phi_H + i \psi_H$. Along the real axis, $\psi_H$ and $\phi_H$ are related by the Hilbert transform $\mathcal{H}$ so that
\begin{equation}
    \phi_H = \mathcal{H} \big(\psi_H\big) \, .
\end{equation}
This, combined with one half of the Cauchy-Riemann equations,
\begin{equation}
    \d_{\mathcal{B}}\psi_H  = -\d_\A \phi_H \, , 
\end{equation}
yields
\begin{equation}
    \d_\beta \psi_H \Big|_{\beta=0}= -\gamma \d_\A \mathcal{H}\big(\psi_H\big) \Big|_{\beta=0}= \gamma \d_\A \mathcal{H}\big(\Xi(\A,T)\Big) \, .
\end{equation}
Therefore, at the surface, to fourth-order,
\begin{multline}
    \langle x_\tau \rangle\Big|_{\beta=0} = -\d_\beta \big(\psi_P + \psi_H\big)\Big|_{\beta=0}  =\epsilon^2 |\F|^2 + \epsilon^2\gamma \bigg(\frac{1}{2}i \big(\F_\A \F^* - \F \F_\A^*\big) - \frac{1}{2} i\big(\F_T\F^* - \F \F^*_T\big) \\+\frac{1}{2}\mathcal{H}\big(|\F|^2_\A\big)\bigg) 
    + 2\epsilon^4 |\F|^4 + \epsilon^2 \gamma^2 \bigg( \frac{1}{2}|\F_\A|^2  - \frac{1}{4}|\F|^2_{\A\A}+ \frac{1}{4}i\d_\A \mathcal{H}\big(\F_T\F^* - \F \F_T^*\big)\bigg)
\end{multline}

\section{Fourth-order mean Lagrangian drift for two waves} \label{twowavesection}

It is natural to ask whether or not an enhanced mean Lagrangian drift can be found from the simplest wave interaction case, that of two irrotational waves with arbitrary wavenumbers $k_1$ and $k_2$ traveling in the same direction. This system is well studied, with nonlinear wave-wave interactions inducing finite amplitude phase speed corrections to each wave while leaving the wave amplitudes time independent \citep{lhPhillips62}.
Without loss of generality we set $k_1 < k_2$ to ensure that solutions decay with depth. Using $\epsilon_1$ and $\epsilon_2$ to represent the small steepness parameters of each wave, solutions valid to second order in each small parameter are given in \cite{Pierson1961}
\begin{multline}\label{piersonx}
    x(\alpha,\beta,\tau) = \alpha + \epsilon_1^2 c_1 e^{2 k_1 \beta}\tau + \epsilon_2^2 c_2 e^{2 k_2 \beta}\tau  -\frac{\epsilon_1}{k_1} e^{k_1\beta} \sin(\theta_1) - \frac{\epsilon_2}{k_2} e^{k_2 \beta} \sin(\theta_2) \\ + \frac{\epsilon_1 \epsilon_2}{g k_1 k_2} (\omega_2 + \omega_1) \omega_2 e^{(k_2 - k_1)\beta} \sin(\theta_2 - \theta_1) \\ - \frac{\epsilon_1 \epsilon_2}{g k_1 k_2} \Big( \frac{\omega_1^3 + \omega_2^3}{\omega_2 - \omega_1}\Big) e^{(k_2 +k_1)\beta}\sin(\theta_2 - \theta_1) \, ,
\end{multline}

\begin{multline}\label{piersony}
    y(\alpha,\beta,\tau) = \beta + \frac{\epsilon_1^2}{2 k_1}e^{2 k_1 \beta} + \frac{\epsilon_2^2}{2 k_2}e^{2 k_2 \beta} + \frac{\epsilon_1}{k_1} e^{k_1 \beta} \cos(\theta_1) + \frac{\epsilon_2}{k_2} e^{k_2 \beta} \cos(\theta_2) \\ +
    \frac{\epsilon_1 \epsilon_2}{g k_1 k_2}(\omega_1^2 + \omega_1 \omega_2 + \omega_2^2) e^{(k_1 + k_2)\beta}\cos(\theta_2 - \theta_1) \\ - \frac{\epsilon_1 \epsilon_2}{g k_1 k_2}\omega_2 (\omega_2 + \omega_1)e^{(k_2-k_1)\beta}\cos(\theta_2 - \theta_1) \, ,
\end{multline}

\begin{multline}\label{piersonp}
    p(\alpha,\beta,\tau) = -g \beta  + 2 \epsilon_1 \epsilon_2 c_1 c_2 e^{(k_2 - k_1)\beta}\cos(\theta_2 - \theta_1) \\ - 2 \epsilon_1 \epsilon_2 c_1 c_2 e^{(k_1 + k_2)\beta}\cos(\theta_2 - \theta_1) \, ,
\end{multline}
where to this order $\omega_1 = \sqrt{g k_1}$, $\omega_2 = \sqrt{g k_2}$, with phases
\begin{equation}
    \theta_1 = k_1 (\alpha - c_1 \tau) \, , \qquad \theta_2 = k_2 (\alpha - c_2 \tau) \, ,
 \end{equation}
where $c_1 = \omega_1/k_1$ and $c_2 = \omega_2/k_2$. As we saw previously for linear theory, the mean Lagrangian drift to this order is simply additive for each wave, and we have
\begin{equation}
    U(\beta) = \epsilon_1^2 c_1 e^{2 k_1 \beta} + \epsilon_2^2 c_2 e^{2 k_2 \beta}\, .
\end{equation}
Solutions to second order can only constrain the Lagrangian pseudomomentum to third order, which has no mean terms. Therefore, we must extend these results to third order in $\epsilon_1$ and $\epsilon_2$ to constrain the pseudomomentum, and as a result the mean Lagrangian drift, to fourth order. Similar to what is necessary for monochromatic waves \citep{Clamond2007}, this requires Doppler shifting the phase by the mean Lagrangian drift as well as allowing for higher order corrections to the linear phase speeds. The modified phase can be expressed as
\begin{equation}\label{phase1}
    \theta_1 = k_1 (\alpha - (c_1 - U(\beta))\tau) \, , \qquad c_1 = \sqrt{\frac{g}{k_1}} + \epsilon_1^2 c_{11} + \epsilon_2^2 c_{12} \, ,
\end{equation}
\begin{equation}\label{phase2}
    \theta_2 = k_2 (\alpha - (c_2 - U(\beta))\tau) \, , \qquad c_2 = \sqrt{\frac{g}{k_2}} + \epsilon_1^2 c_{21} + \epsilon_2^2 c_{22} \, ,
\end{equation}
where $c_{11}$, $c_{12}$, $c_{21}$ and $c_{22}$ are to be determined corrections \citep[there are no phase speed corrections proportional to $\epsilon_1 \epsilon_2$,][]{lhPhillips62}. From the second order solutions \eqref{piersonx}--\eqref{piersonp}, third order terms are added of the form, e.g. for $x$,
\begin{equation}
    \sum_{i + j = 3}\epsilon_1^i \epsilon_2^j x^{(i,j)}(\alpha,\beta,\tau) \, ,
\end{equation}
and likewise for $y$ and $p$. The third order expansions, using the expanded phases \eqref{phase1}--\eqref{phase2}, are then inserted into the Euler equations \eqref{xEuler}--\eqref{yEuler}, irrotational condition \eqref{irrotational} and Jacobian gauge ($\J=1$) \eqref{Jacobian}. Terms proportional to the same powers of $\epsilon_1$ and $\epsilon_2$ are then matched. Because of our phase expansion, all third order terms are nonsecular and can be found by assuming sinusoidal $\alpha$ dependence and exponential $\beta$ dependence. In the process the corrections to the phase speeds are also constrained. These corrections were found to be
\begin{align}\label{phasecorrections1}
    c_1 &= \sqrt{\frac{g}{k_1}}\Big(1 + \tfrac{1}{2}\epsilon_1^2\Big) + \epsilon_2^2 \frac{k_1}{k_2}\sqrt{\frac{g }{k_2}}\, , \\
    c_2 &= \sqrt{\frac{g}{k_2}}\Big(1 + \tfrac{1}{2}\epsilon_2^2\Big) + \epsilon_1^2 \sqrt{\frac{g}{k_1}} \, ,\label{phasecorrections2}
\end{align}
which match exactly with the results of \cite{lhPhillips62} derived by a purely Eulerian approach. In addition to the standard finite amplitude Stokes correction to the phase speeds, there also exist corrections due to tertiary nonlinear interactions. Notably, these corrections are not symmetric; the relative increase of $c_1$ from wave 2 depends on the ratio of $k_1/k_2$ whereas the relative increase of $c_2$ from wave 1 solely depends on wave 1. 

From these third order solutions, including corrections to the phase speeds, we can constrain the fourth order mean Lagrangian drift $\langle x_\tau \rangle$ from the mean pseudomomentum at each vertical level. This yields
\begin{multline}
    \langle x_\tau \rangle = \epsilon_1^2c_1e^{2 k_1\beta}  + 2\epsilon_1^4 e^{4 k_1 \beta}\sqrt{\frac{g}{k_1}} + \epsilon_2^2c_2 e^{2 k_2\beta} + 2\epsilon_2^4 e^{4 k_2 \beta}\sqrt{\frac{g}{k_2}} +\\\epsilon_1^2 \epsilon_2^2 \bigg( \frac{\sqrt{g}(\sqrt{k_1}-\sqrt{k_2})^2 (\sqrt{k_1}+\sqrt{k_2})^3}{k_1^2 k_2}e^{2(k_2-k_1)\beta} + \frac{(\sqrt{gk_1}+\sqrt{g k_2})(k_1+k_2)^3}{k_1^2 k_2^2}e^{2(k_1 + k_2)\beta} \\- 2\frac{ k_2(\sqrt{g k_1} + \sqrt{g k_2})}{k_1^2}e^{2 k_2 \beta}\bigg) \, , \label{2wavedrift}
\end{multline}
where $c_1$ and $c_2$ include nonlinear corrections \eqref{phasecorrections1}--\eqref{phasecorrections2}. Reassuringly, the ``monochromatic terms" for each wave (i.e. those not multiplied by $\epsilon_1^2 \epsilon_2^2$) exactly match the results found above for purely monochromatic waves \eqref{monodriftClamond}. However, there are also interesting interaction terms which are best interpreted through certain limiting examples.

In the case where $k_2 \gg k_1$, such that we are considering the dynamics of a short wavelength wave riding atop a long wavelength wave, we see from \eqref{phasecorrections1} that the modification to the phase speed of the long wave from the short wave is negligible. However, the phase speed of the short wave is modified by the long wave, precisely by the surface mean Lagrangian drift of the long wave, indicating that the short wave is mainly advected by the surface mean Lagrangian drift of the long wave, which varies much more slowly with depth and as such acts as a constant external current. The mean Lagrangian drift for this limit can be written,
\begin{equation}
    \langle x_\tau \rangle = \epsilon_1^2c_1e^{2 k_1\beta}  + 2\epsilon_1^4 e^{4 k_1 \beta}\sqrt{\frac{g}{k_1}} + \epsilon_2^2c_2 e^{2 k_2\beta} + 2\epsilon_2^4 e^{4 k_2 \beta}\sqrt{\frac{g}{k_2}} \, ,
\end{equation}
where, interestingly, the interaction terms not connected to shifts in the phase speed vanish in this limit, which matches very well with the idea that nonlinear wave-wave interactions are strongest between waves of similar wavenumbers \citep{hasselmann1962}. Therefore, the total mean Lagrangian drift in this long-short wave system can be expressed as a sum of an unmodified long wave drift and a short wave drift whose phase speed is Doppler shifted by the surface drift of the long wave.

Next we can consider the limit where $k_2 = k_1 + \delta k$, where $\delta k$ is small and positive, so that the wave field appears as a series of groups. To lowest order, $k_1/k_2 \approx 1$ and the Doppler shift of each wave is symmetric, with the drift given by
\begin{multline}
    \langle x_\tau \rangle = \epsilon_1^2\big(1 + \tfrac{1}{2}\epsilon_1^2)  e^{2 k_1 \beta}\sqrt{\frac{g}{k_1}} +\epsilon_2^2\big(1 + \tfrac{1}{2}\epsilon_2^2)  e^{2 k_1 \beta}\sqrt{\frac{g}{k_1}}+ 2( \epsilon_1^4 +\epsilon_2^4) e^{4 k_1 \beta}\sqrt{\frac{g}{k_1}}  \\+\epsilon_1^2 \epsilon_2^2 \Big(16 e^{4k_1\beta} - 2 e^{2 k_1 \beta}\Big)\sqrt{\frac{g}{k_1}} + O(\delta k ) \, .
\end{multline}
Here we see that in addition to the standard fourth order monochromatic drift of each wave, there is a fourth order interaction term which is positive near the surface, zero at $k_1\beta > \tfrac{1}{2}\ln(1/4)$, and negative below, analogous to what was found for wave packets via a narrow-banded approach in $\S$\ref{packets}.

In both of these limits, the two waves interact to produce fourth order positive corrections to the near surface mean Lagrangian drift, which match the results presented in this paper. However, given the analytical difficulty of deriving a higher-order mean Lagrangian drift for even this relatively simple system, it is unlikely that such a direct approach will prove useful for more complex wave configurations.

\bibliographystyle{jfm}

\bibliography{ref}

\end{document}